\documentclass{aa}  

\usepackage{multirow}
\usepackage{xcolor}
\usepackage{graphicx}
\usepackage{rotating} 
\usepackage{comment}
\usepackage[para,online,flushleft]{threeparttable}
\usepackage{caption}
\usepackage{subcaption}
\usepackage{adjustbox}
\usepackage{gensymb}
\usepackage{overpic} 
\usepackage{natbib,twoopt}
\usepackage[hyphenbreaks]{breakurl}
\usepackage[breaklinks]{hyperref}      
\bibpunct{(}{)}{;}{a}{}{,}             
\definecolor{cobalt}{rgb}{0.06, 0.2, 0.65}
\hypersetup{
  colorlinks,
  citecolor=cobalt,
  linkcolor=[rgb]{0.8, 0.2, 1.0},
  urlcolor=cobalt,
}
\makeatletter
  \newcommandtwoopt{\citeads}[3][][]{\href{http://adsabs.harvard.edu/abs/#3}%
    {\def\hyper@linkstart##1##2{}%
     \let\hyper@linkend\@empty\citealp[#1][#2]{#3}}}
  \newcommandtwoopt{\citepads}[3][][]{\href{http://adsabs.harvard.edu/abs/#3}%
    {\def\hyper@linkstart##1##2{}%
     \let\hyper@linkend\@empty\citep[#1][#2]{#3}}}
  \newcommandtwoopt{\citetads}[3][][]{\href{http://adsabs.harvard.edu/abs/#3}%
    {\def\hyper@linkstart##1##2{}%
     \let\hyper@linkend\@empty\citet[#1][#2]{#3}}}
  \newcommandtwoopt{\citeyearads}[3][][]%
    {\href{http://adsabs.harvard.edu/abs/#3}
    {\def\hyper@linkstart##1##2{}%
     \let\hyper@linkend\@empty\citeyear[#1][#2]{#3}}}
\makeatother

\def\gs{\mathrel{\raise0.35ex\hbox{$\scriptstyle >$}\kern-0.6em \lower0.40ex\hbox{{$\scriptstyle \sim$}}}}
\def\ls{\mathrel{\raise0.35ex\hbox{$\scriptstyle <$}\kern-0.6em \lower0.40ex\hbox{{$\scriptstyle \sim$}}}}
\usepackage{txfonts}
\usepackage{rotating}     
\usepackage{adjustbox}    
\usepackage{subcaption}      
\usepackage{booktabs}        
\usepackage{caption}         

\usepackage{hyperref}
\usepackage{orcidlink}
\begin{document}

   \title{The JWST Emission Line Survey: Multi-wavelength properties of Paschen line emitters at cosmic noon}
   
   \titlerunning{Multi-wavelength properties of Paschen line emitters at cosmic noon}
   \author{Malte Brinch\orcidlink{0000-0002-0245-6365}\inst{1,2}
   \and
   Edo Ibar\orcidlink{0009-0008-9801-2224}\inst{1,2}
   \and
   Luis Ossa-Fuentes\orcidlink{0009-0002-3124-1328}\inst{1,2}
   \and
   Corey Pirie\orcidlink{0009-0003-5303-6920}\inst{3}
   \and
   Philip Best\orcidlink{0000-0001-5081-4801}\inst{3}
   \and
   R. K. Cochrane\orcidlink{0000-0001-8855-6107}\inst{4}
   \and
   Kenneth Duncan\orcidlink{0000-0001-6889-8388}\inst{3}
   \and
   James Dunlop\orcidlink{0000-0002-1404-5950}\inst{3}
   \and 
   Sophia Flury\orcidlink{0000-0002-0159-2613}\inst{3}
   \and
   Catherine Hale\orcidlink{0000-0001-6279-4772}\inst{3}
   \and
   Jorryt Matthee\orcidlink{0000-0003-2871-127X}\inst{5}
   \and
   Derek McLeod\orcidlink{0000-0003-4368-3326}\inst{3}
   \and
   Joshua Selfridge\orcidlink{0000-0001-5575-7906}\inst{4}
   \and
   David Sobral\orcidlink{0000-0001-8823-4845}\inst{6}
   \and
   Harry Stephenson\orcidlink{0000-0002-0777-1591}\inst{7}
   \and
   J. P. Stott\orcidlink{0000-0002-1679-9983}\inst{7}
   \and
   A.M. Swinbank\inst{8}
   }

\institute{Instituto de Física y Astronomía, Universidad de Valparaíso, Avda. Gran Breta\~{n}a 1111, Valparaíso, Chile\\
              \email{malte.brinch@uv.cl}
    \and
    Millennium Nucleus for Galaxies (MINGAL), Chile
    \and
    Institute for Astronomy, University of Edinburgh, Royal Observatory, Edinburgh, EH9 3HJ, UK
    \and
    Jodrell Bank Centre for Astrophysics, University of Manchester, Oxford Road, Manchester M13 9PL, UK
    \and
    Institute of Science and Technology Austria (ISTA), Am Campus 1 3400, Klosterneuburg, Austria
    \and
    BNP Paribas Corporate \& Institutional Banking, Torre Ocidente Rua Galileu Galilei, 1500-392 Lisbon, Portugal
    \and 
    Department of Physics, Lancaster University, Lancaster, United Kingdom, LA1 4YB
    \and
    Centre for Extragalactic Astronomy, Department of Physics, Durham University, South Road, Durham DH1 3LE
             }

   \date{Received 8 April 2026 / Accepted 8 September 2026}  
   
 
  \abstract
{{\it Context.} 
Narrowband JWST/NIRCam selection isolates star-forming galaxies in well-defined redshift slices, whereas ALMA and archival far-IR/submillimeter data reveal the obscured component of their growth from dust-enshrouded star formation to cold gas. We combined these views to measure Paschen-line star formation rates (SFRs), infer nebular line extinction from EAZY-derived $A_{\rm V}$, and probe the average FIR-to-submillimeter emission through weighted-median stacking.
Through this work, we aim to place constraints on galaxy growth at cosmic noon.
\\
{\it Aims.} We characterize the stellar, nebular, and far-IR properties of Paschen-line selected galaxies in the COSMOS field, focusing on two ensembles at $z\sim1.50$ (Paschen-$\alpha$) and $z\sim2.65$ (Paschen-$\beta$). We quantify short- versus long-timescale star formation and dust attenuation, and we constrain obscured star formation and gas masses.
\\
{\it Methods.} We selected robust narrowband excess samples from the JWST Emission Line Survey (JELS) F466N/F470N mosaic (Paschen-$\alpha$: 77 sources, Paschen-$\beta$: 31 sources), removed active galactic nuclei via X-ray and radio crossmatching, and assembled multi-wavelength photometry (JWST/NIRCam+MIRI, HST, Spitzer, Herschel, SCUBA-2, AzTEC, ALMA). Photometric redshifts and galaxy stellar properties were derived with EAZY and STARDUST. 
We estimated line SFRs using recent Paschen-line calibrations, with nebular line extinction inferred from EAZY-derived $A_{\rm V}$. We explored their far-IR to submillimeter properties through weighted-median image stacking.
\\
{\it Results.} EAZY fits indicate a wide range of stellar masses ($\rm \log(M_\star/M_\odot)\sim7.4$–$10.3$) and modest SFRs ($\rm SFR_{\rm 100\,Myr}\sim0.1-27\,\rm M_{\odot}\,yr^{-1}$). We derived line SFRs of $\rm 0.1-22.0\, M_{\odot}\, yr^{-1}$ and $\rm 0.7-47.9\, M_{\odot}\, yr^{-1}$ for Pa-$\alpha$ and Pa-$\beta$, respectively.
Most sources appear to be experiencing short-term bursts, with line-based SFRs exceeding the 100 Myr spectral energy distribution (SED) averages. There is a weakly decreasing trend in burstiness with increasing stellar mass. Nebular attenuation correlates positively with stellar mass and SFR. From the stacking analysis of the available images, we find no $>3\sigma$ stacked detections at wavelengths longer than MIRI.  
From fitting the stellar component of the weighted median SED, we obtained ${\rm SFR}_{\rm UV,med}=\rm 0.48\pm0.11\, M_{\odot}\, yr^{-1}$ and ${\rm SFR}_{\rm UV,med}=\rm 3.92\pm0.47\, M_{\odot}\, yr^{-1}$ for Pa-$\alpha$ and Pa-$\beta$ samples, respectively.
ALMA Band 4 yields the tightest constraints for the inferred obscured SFR and gas mass, for which the median $3\sigma$ upper limits are $\rm SFR_{IR}\lesssim53\,M_\odot\,yr^{-1}$ (Pa-$\alpha$) and $\rm \lesssim44\,M_\odot\,yr^{-1}$ (Pa-$\beta$). 
For the gas-mass bounds, we obtained $\rm\lesssim10^{11.2}\,M_\odot$ (Pa-$\alpha$) and $\rm\lesssim10^{10.6}\,M_\odot$ (Pa-$\alpha$) under standard assumptions.
\\
{\it Conclusions.} Paschen-selected samples reveal widespread short-timescale star formation across a wide mass range, but current far-IR/submillimeter data only provide upper limits on obscured SFR and interstellar medium mass. Spectroscopic metallicity diagnostics and deeper ALMA continuum observations are required to robustly measure dust-obscured star formation and gas reservoirs in these line-emitting samples.}


   \maketitle
%

\section{Introduction}
A central goal of galaxy evolution studies is to understand how rapidly galaxies build their stellar mass, how much of that growth is hidden by dust, and how that obscured growth is connected to the available cold-gas reservoir. How star formation is distributed across mass, environment, and cosmic time also needs to be understood.
These questions are particularly important at cosmic noon ($z\sim1-3$), which is when the cosmic star-formation rate (SFR) density \citep{Dickinson2014} peaks and dust-obscured star formation becomes dominant. In this regime, the key unknowns are not only the total SFR but also the relative contribution of short-timescale bursts, the degree of nebular attenuation, and whether the gas reservoirs of these galaxies are large enough to sustain the observed activity. 
To obtain these critical measurements, it is essential to select unbiased samples of star-forming galaxies across cosmic time. To determine key physical properties such as SFR, stellar mass, and gas content, galaxies can be selected through spectroscopic redshift surveys. 
Their physical properties can be traced through complementary windows: ultraviolet (UV) light from young stars, optical lines from ionized gas, far-infrared (FIR) dust emission, and radio continuum from obscured star formation \citep[e.g.,][]{Kennicutt1998, Hopkins2006, Kennicutt2012}. 
While each tracer comes with systematic uncertainties, taken together they have yielded a relatively coherent picture of galaxy growth out to cosmic noon at $z\sim6$ \citep{Dickinson2014, Barrufet2025}.
However, at cosmic noon, where the cosmic star formation peaks, roughly 85\% of the total SFR density is obscured by dust \citep[e.g.,][]{Ibar2013, Casey2014, Dunlop2017, Thompson2017, Dudzeviciute2020, Traina2024}.
As a result, UV-selected surveys miss much of the star formation at these epochs, which biases our understanding of galaxy growth.
This also obscures the physical link between star formation and the interstellar medium. Observations in the submillimeter and FIR have revealed a fundamentally different view of galaxy demographics and physical conditions, one where heavily dust-enshrouded systems dominate the energy budget at least out to $z\sim6$ \citep{Casey2014, Fudamoto2020, Khusanova2021, Fudamoto2021, Zavala2021, Algera2023}.

Studies leveraging the Atacama Large Millimeter Array (ALMA) sensitivity have shown that the spatial distribution of dust-obscured star formation in massive galaxies is often compact and clumpy \citep{Hodge2016} and that the dust continuum correlates strongly with molecular gas mass \citep{Scoville2016, Scoville2017}, providing a powerful means to trace the fuel for star formation. 
Early Herschel stacking studies \citep{Magdis2012, Santini2014, Bethermin2015} already provided constraints on average dust emission, gas fractions, and depletion times in star-forming galaxies, which were later extended with Institut de radioastronomie millimétrique (IRAM) and ALMA studies \citep[e.g.,][]{Genzel2015, Tacconi2018, Wang2024, Aravena2024} to higher redshift and larger samples. This has provided key insight into how galaxies build their stellar mass over cosmic time. 

At the same time, the James Webb Space Telescope (JWST) Near Infrared Camera (NIRCam), with its suite of narrowband filters and unprecedented sensitivity, provides a complementary selection method that identifies line-emitting galaxies in narrow redshift slices using narrowband surveys \citep{Duncan2025}. This narrowband technique enables the clean selection of star-forming galaxies across cosmic time. Studies such as HiZELS have shown the ability of narrowband surveys to target faint H$\alpha$ emitters out to $z\sim2$ \citep{Sobral2013} using ground-based narrowbands, and with JWST, this can be pushed to fainter lines and higher redshifts. The deep JWST narrowband selection allows for the study of lower-mass line emitters with lines other than H$\alpha$, which is typically used in the literature \citep{Cochrane2018}, that are less affected by dust obscuration. This allowed us to probe the properties of a hitherto poorly explored galaxy population relevant for the overall evolution scenario of galaxies, as they are more numerous than the massive galaxies selected before the advent of JWST. 

In this paper, we combine ALMA’s ability to constrain dust-obscured star formation and cold molecular gas with JWST’s clean selection of narrowband line emitters from the JWST Emission Line Survey (JELS). 
We analyze a sample of Paschen-$\alpha$ and Paschen-$\beta$ emitters, which provides a powerful way to probe diverse galaxy populations, from dusty starbursts to low-mass UV-bright emitters. We aim to constrain the fundamental properties of the Paschen line emitters, such as stellar mass, SFR, and extinction ($A_V$, $A_{\rm Line}$), as well as their median dust and gas properties through stacking analysis.
This work investigates three related questions: {\it (i)} whether these Paschen-line selected galaxies form stars in short-lived bursts or in a steadier mode, {\it (ii)} how much of their star formation is hidden by dust, and {\it (iii)} whether they have substantial cold-gas reservoirs.

The paper is organized as follows. In Sect.~\ref{sec:data} we describe the JELS narrowband observations and the ancillary datasets used in this study, and 
we detail our narrowband excess selection, spectroscopic crossmatches, and active galactic nucleus (AGN) rejection. Section~\ref{sec:methods} presents our analysis methods: spectral energy distribution (SED) fitting, the emission-line SFR and extinction prescriptions, and the weighted-median stacking procedure, including our multiple-imputation plus cluster-bootstrap scheme. In Sect.~\ref{sec:Results} we report the derived galaxy properties, quantify short- versus long-timescale SFRs, and describe the limits on obscured SFR and gas mass. In Sect.~\ref{sec:Discussion} we discuss the implications of the results, including the tentative evidence of massive-galaxy rejuvenation. Finally, Sect.~\ref{sec:Conclusions} provides a summary of our conclusions.

Throughout the paper, we adopt a standard $\Lambda$ cold dark matter cosmology
with $H_{\rm 0}=70\,{\rm km\,s^{-1}\,Mpc^{-1}}$, $\Omega_{\rm m} = 0.3$, 
and $\Omega_{\rm \Lambda} = 0.7$. All magnitudes are expressed in the
AB system \citep{Oke1974}. We adopted a \cite{Chabrier2003} stellar initial mass
function. The results are reported with a 68\% confidence interval uncertainty.
\section{Selection and data}\label{sec:data}

\subsection{JELS}
The JWST Emission Line Survey (GO $\#2321$, PI: Best) is detailed comprehensively in \cite{Duncan2025}. Briefly, the survey employs the JWST/NIRCam long-wavelength narrowband filters F466N and F470N. For a description of the filter properties and their transmission curves, refer to Table 1 and Fig. 2 in \cite{Pirie2024}.
Parallel observations were conducted in the short-wavelength channel using the F212N filter, which has a pivot wavelength of $\lambda_{\rm pivot} = 2.1213\,\mu\rm m$ and a bandwidth of $\Delta\lambda = 0.0274\,\mu\rm m$, along with the F200W filter. However, these data are not analyzed in this paper \citep[see discussion in ][]{Duncan2025}.
The JELS survey employed a $3 \times 3$ mosaic design with 57\% overlap between adjacent columns, using the "Medium8" observing strategy. This involved nine groups for the F466N filter and ten groups for the F470N filter, resulting in approximately 1000 seconds of on-sky exposure time per observation. To address bad pixels and cosmic rays, a three-point intra-module dithering pattern was employed, including two sub-pixel dithers at each position.
This configuration provided continuous coverage of a 63 arcmin$^2$ region within the Cosmic Evolution Survey \citep[COSMOS;][]{Scoville2007} field, centered at coordinates (RA, Dec) = (150.125, 2.333) degrees. The total on-sky integration time was around 6 ks for the full mosaic, with double-depth imaging (12 ks) over the central $\sim40\%$ of the field. Overall, the program required 43.0 hours of observing time.
The $5\sigma$ limiting emission-line fluxes are $\rm 10^{-18.04}-10^{-17.88}\, erg/s/cm^{2}$ for F466N and $\rm 10^{-18.12}-10^{-17.92}\, erg/s/cm^{2}$ for F470N filters, respectively, depending on if the deepest 35\% or 90\% of the field is targetted \citep{Duncan2025}. The line fluxes are up to about two times fainter than current literature slitless spectroscopic surveys \citep{Duncan2025}. 

Relevant for our study, JELS targets the following two lines:
\begin{itemize}
    \item Paschen-$\alpha$ (Pa-$\alpha$): This line arises from the hydrogen $n=4-3$ electron transition and is emitted at $\lambda_{\rm P\alpha}=1.87561\,\rm \mu m$. Using the F466N and F470N narrowband bandwidths, the line falls within the narrowbands at $z=1.4676-1.4964$ and $z=1.4967-1.5240$, respectively.
    \item Paschen-$\beta$ (Pa-$\beta$): This line arises from the hydrogen $n=5-3$ electron transition and is emitted at $\lambda_{\rm P\beta}=1.28216\,\rm \mu m$. Using the F466N and F470N narrowband bandwidths, the line falls within the narrowbands at $z=2.6088-2.6509$ and $z=2.6520-2.6918$, respectively.
\end{itemize}
Pa-$\alpha$ and Pa-$\beta$ are recombination lines powered by Lyman-continuum photons from O and early B stars with lifespans of only $\rm\sim3-10\, Myr$. Similar to H$\alpha$, they trace massive star formation over short timescales of $\rm\sim10\, Myr$, but with lower dust extinction \citep[$\gtrsim5\times$ lower;][]{Calzetti2007}. 
This makes them useful for probing recent massive star formation with much less sensitivity to dust than UV tracers \citep{Leitherer1999, Rieke2009, Kennicutt2012, Flores2021} and for constraining the cosmic SFR density (Ossa-Fuentes et al. submitted to A\&A).

We note that the images used in this paper are the newer version of the JELS imaging products, referred to as
v1.0 images, compared to the v0.8 images used in \citet{Pirie2024} and \citet{Stephenson2025}. These images incorporate re-observations to better mitigate scattered-light issues and use an updated version of the JWST pipeline for image reduction. We refer the reader to Appendix A of \cite{Duncan2025} for details on the updated v1.0 images.
\subsection{JWST and HST data}
To distinguish between the different emission lines that the narrowbands can target (e.g., PAHs, H$\alpha$, Pa-$\alpha$, Pa-$\beta$, O[III]), we required ancillary data for SED fitting to obtain their photometric-redshift probability distributions.
We made use of the Public Release IMaging for Extragalactic Research \citep[PRIMER;][]{Dunlop2021}, a JWST Cycle 1 Treasury Program, large-area, homogenous, deep JWST/NIRCam and Mid-Infrared Instrument (MIRI) imaging survey. The PRIMER survey consists of eight NIRCam (F090W, F115W, F150W, F200W, F277W, F356W, F410M, F444W) and two MIRI (F770W, F1800W) bands. JELS is embedded within the PRIMER footprint. This allowed us to exploit the deep JWST photometry to analyze our narrowband excess sources in a broader spectral context.
Given that MIRI coverage in PRIMER is incomplete, not all of our sources have MIRI data. We show the number of sources with available data in each band in Table \ref{tab:JWSTflux}.

The COSMOS field is also covered by a wide range of Hubble Space Telescope (HST) photometry, which we make use of as part of the SED fitting of the narrowband excess source, including Advanced Camera for Surveys (ACS) (F435W, F606W, F606W) and Wide Field Camera 3 (WFC3) (F275W, F125W, F140W, F160W). The use of these bands helps constrain the redshift probability distribution and the physical parameters that are related to the stellar component of the SED. 
Note that, especially for the F275W and F435W bands, coverage is missing for some sources. We show the number of sources with available data in each band in Table \ref{tab:Hubbleflux}.

Following \cite{Pirie2024}, we use 300 mas-diameter aperture fluxes to detect narrowband-excess line emitters via color selection and to maximize the signal-to-noise ratio (S/N; see Sect. \ref{sec:Nline sel}). 
For further analysis, we use 600 mas diameter aperture photometry performed on images homogenized to a common point spread function (PSF) from the JELS catalog \citep{Pirie2024}. The 600 mas diameter aperture measurement was
chosen to increase the light collection area when calculating the source
line fluxes and obtain more accurate physical parameter estimates when performing the SED fitting.
For the MIRI data, we performed aperture photometry using 600 mas apertures. 

To account for missing flux when using a 600 mas aperture and following \cite{McLeod2024, Shuntov2025, Liu2026}, we calculate the aperture correction by taking the ratio between the FLUX\_AUTO value \citep{Kron1980} from Source Extractor and the 600 mas aperture flux using the F444W band and add a further $10\%$ to account for light outside of the Kron aperture.
For each source, we scale the fluxes and associated uncertainties by this correction.
\subsection{Spectroscopic redshift selection}
To obtain the available spectroscopic redshifts for the line-emitter galaxies, we use three catalogs. Firstly, the catalog used for the Cosmic Assembly Near-infrared Deep Extragalactic Legacy Survey (CANDELS) by \cite{Kodra2023} \citep[see also][]{Grogin2011, Koekemoer2011}. Secondly, the DR1.1 release of the COSMOS Spectroscopic Redshift Compilation \citep{Khostovan2025} with confidence levels at $\geq95\%$, which only includes reliable and very reliable redshifts with multiple high S/N line detections. Thirdly, we use the public JWST Near Infrared Spectrograph (NIRSpec) v4.4 release of the DAWN JWST Archive \citep{Heintz2025}\footnote{\url{https://dawn-cph.github.io/dja/index.html}} with redshift inspection grade three, classified as robust redshifts. If spectroscopic redshifts are available when performing SED fitting, galaxy properties are estimated by fixing the galaxy redshift to the spectroscopic redshift.
\subsection{Narrowband line-emitter selection}\label{sec:Nline sel}
Galaxies were selected according to a number of criteria. First, the narrowband excess sources must fulfill color excess and excess significance criteria as given by Eq. 2 in \cite{Pirie2024}:
\begin{equation}
    \Sigma=\frac{1-10^{-0.4(BB-NB)}}{10^{-0.4(ZP-NB)}\sqrt{\sigma_{NB}^2+\sigma_{BB}^2}},
\end{equation}
where either the narrow-narrow (N-N) or broad-narrow (B-N) band conditions are fulfilled, with F466N and F470N being the narrowband and F444W being the broadband. We note that $ BB - NB$ is the narrow-to-broadband color calculated from the band magnitudes.
For narrow-narrow band excesses, the BB was replaced with NB$_{1}$ and NB$_{2}$. 
The term ZP is the zero point of the narrowband, which was set to 23.9 mag. The terms $\sigma_{NB}$ and $\sigma_{BB}$ are photometric flux density errors (in $\mu$Jy) for the NB and BB filters, respectively, for each source. 
The specific conditions are shown in Table \ref{tab:colorcond}. 
The narrowband excess significance parameter separates true narrowband excess sources from scatter due to photometric uncertainties. In contrast, the color excess cut acts as an observed equivalent width cut and is required due to systematic scatter even at bright magnitudes \citep[see Fig. 8 in][]{Pirie2024}.
Second, we adopt a conservative selection approach where the source has a well-constrained photometric redshift probability distribution. 
The peak and median values of the distribution must lie within the redshift range where the emission lines would fall within the narrowband. Third, if a source has a spectroscopic redshift that falls within the narrowband redshift ranges, then it automatically passes these photometric redshift criteria. 
For the SFR calculation, we adopted the line flux from whichever estimate (N-N or B-N) has the highest S/N, following Eqs. 3 and 6 in \cite{Pirie2024}.
\begin{table}[h!]
\caption{Color conditions for narrowband excess sources using 300 milliarcsecond diameter apertures with the broadband (F444W) and narrowband (F466N, F470N).}
\label{tab:colorcond}
\centering
\small
\begin{tabular}{lcc}
\hline\hline
Selection & Color excess (dex) & Excess significance ($\Sigma$) \\
\hline
Broad--Narrow & $>0.30$ & $>3.0$ \\
Narrow--Narrow & $>0.15$ & $>2.5$ \\
\hline
\end{tabular}
\tablefoot{The excess significance was calculated using Eq. 2 of \citet{Pirie2024}. The small aperture size was only used to maximize the S/N of a detection, not for flux measurements.}
\end{table}

The JWST/NIRCam+MIRI images were also visually inspected for each source to investigate potential false detections. The F356W band was paid special attention to, as it is the deepest in the PRIMER survey, as well as the F444W, F466N, and F470N bands, as they are crucial for classifying galaxies as narrowband line emitters. 
We did not find any galaxies that were to be removed from this inspection. This is in contrast to \cite{Pirie2024}, which removed $\sim50\%$ of the galaxies from their H$\alpha$ emitter sample at $z=6.1$ using the v0.8 catalog, with tests on the v1.0 catalog only having removed $\sim11\%$. 

\subsection{AGN removal}
To properly characterize the star-forming properties of the galaxies in the Paschen line emitter samples, we needed to remove any galaxies that have a strong contribution to their SED from an AGN.
We removed any sources previously cataloged at X-ray wavelengths by crossmatching our samples with the C-COSMOS Legacy Survey Point Source Catalog \citep{Civano2016} to within the 0.5-arcsecond angular resolution of the Chandra X-ray telescope.
Using the soft band ($0.5-2\,\rm KeV$) limiting depth of $2.2\times10^{-16}\,\rm erg\, cm^{-2}\, s^{-1}$ and assuming a power law Spectral shape with photon index $\Gamma=2$ together with the SFR conversion of \cite{Mineo2012} for point sources and no intrinsic absorption or AGN contribution, the limiting SFR is $\sim2410\,\rm M_{\odot}\,yr^{-1}$ for Pa-$\alpha$ and $\sim9701\,\rm M_{\odot}\,yr^{-1}$ for Pa-$\beta$. Given our estimated SFRs from SED fitting and the Paschen lines, any detection in the X-ray must be from a contributing AGN. 

For our samples, this emission is expected to be produced via inverse Compton scattering by compact and highly ionized regions surrounding an AGN. 
This could remove some sources presenting powerful thermal X-ray emission. However, this does not affect our analysis, as no X-ray emitting sources were identified for the Pa-$\alpha$ and Pa-$\beta$ samples.

We also crossmatched our samples with the Karl G. Jansky Very Large Array (VLA)-COSMOS $3\,\rm GHz$ Large project \citep{Smolcic2017}, AGN counterpart catalog \citep{Delvecchio2017}, and multi-wavelength counterparts catalog \citep{Smolcic2017b} using a 0.8 arcsecond search radius, the same as \cite{Smolcic2017b}.
We find one AGN in the Pa-$\alpha$ sample with a 3GHz flux density of $37.4\pm2.9\,\rm \mu Jy$, which is identified as an AGN from SED decomposition and is classified as a moderate-to-high radiative luminosity AGN (HLAGN) as per \cite{Delvecchio2017}. This source was therefore excluded from the sample. 
We also find one star-forming galaxy with a detection of $11.0\pm2.2\,\rm \mu Jy$ and a peak photometric redshift of $1.5932$, very close to the narrowband. The SFR obtained from the total infrared (IR) luminosity ($8-1000\,\rm \mu m$ rest-frame) is ${\rm SFR}_{\rm IR}=16.42\,\rm M_{\odot}\,yr^{-1}$, assuming the \cite{Kennicutt1998} conversion factor, and scaled to a \cite{Chabrier2003} initial mass function.
For the Pa-$\beta$ sample, we find one galaxy with a radio counterpart. The galaxy is detected with a flux density of $17.0\pm2.4\,\rm \mu Jy$ at $3\,\rm GHz$ and has a spectroscopic redshift of $z=2.6764$ and ${\rm SFR}_{\rm IR}=409.12\,\rm M_{\odot}\,yr^{-1}$. 
The two star-forming galaxies are not classified as an AGN by any of the metrics in \cite{Delvecchio2017}, and are classified as part of their clean star-forming galaxy sample. This means the sources have no $>3\sigma$ radio excess in $\log(L_{\rm 1.4\, GHz}/{\rm SFR}_{\rm IR})$. We therefore chose to keep them in our samples. 

Lastly, we followed the procedure of \cite{Donley2012} using Spitzer/Infrared Array Camera (IRAC) channel 1-4 fluxes from the S-COSMOS survey \citep{Sanders2007} to search for potential AGN. 
It should be noted that the criteria were designed to be highly reliable for luminous AGN, but the method is incomplete for low-luminosity or host-dominated AGN. IRAC-only diagnostics become ineffective at $z\gtrsim2.5$ because the bands start tracing the $1.6\,\rm \mu m$ stellar bump, which can produce AGN-like IRAC colors in ordinary galaxies \citep{Messias2012, Cowley2016, Radcliffe2018}. 
We use the same search radius as \cite{Donley2012} of 2 arcsec, but also check smaller radii of 0.8 and 0.3 arcsec. We find that one Pa-$\alpha$ and one Pa-$\beta$ emitter meet the \cite{Donley2012} criteria for potential AGN and are matched within all three search radii. We remove the Pa-$\alpha$ from the sample.

The Pa-$\beta$ emitter is one of the few galaxies that have JWST/NIRSpec spectra available. The spectrum is      G235M$\_$F170LP from the Blue Jay survey \citep[ID: GO$\#$1810, PI: Sirio Belli;][]{Belli2025} with a spectroscopic redshift of $z_{\rm spec}=2.6371$. 
We compare with the redshift-dependent Baldwin--Phillips--Terlevich (BPT) diagram of \cite{Kewley2013} to see if the galaxy falls within the AGN or Star-forming galaxy regime. We find $\log({\rm [NII]6583/H\alpha})=-0.39$ and $\log({\rm [OIII]5007/H\beta})=0.16$, putting the source in the Star-forming galaxy regime. 
Given the evidence from the NIRSpec data and the reservations about the IRAC selection at $z\gtrsim2.5$, we choose to keep the Pa-$\beta$ emitter in the sample.  
In Table \ref{tab:sourcecount}, we show the final number of sources we have for each emission line and redshift after applying our selection criteria.
\begin{table}[h!]
\caption{Star-forming galaxy counts after applying the selection criteria for narrowband excess sources and the emission lines targeted by each filter.}
\label{tab:sourcecount}
\centering
\begin{tabular}{lccc}
\hline\hline
Line & Redshift & $N_{\rm F466N}$ & $N_{\rm F470N}$ \\
\hline
Pa-$\alpha$ & 1.50 & 19 & 58 \\
Pa-$\beta$  & 2.65 & 18 & 13 \\
\hline
\end{tabular}
\end{table}

\subsection{Mid-IR and FIR data}
We used wide field Spitzer Space Telescope Multiband Imaging Photometer (MIPS) observations taken by the S-COSMOS survey at $24\,\rm \mu m$, $70\,\rm \mu m$, and $160\,\rm \mu m$ (\citealt{Sanders2007}); Herschel Space Observatory Photodetector Array Camera and Spectrometer (PACS) observations taken by the PEP survey \citep{Lutz2011} at $100\,\rm \mu m$ and $160\,\rm \mu m$; Herschel-Spectral and Photometric Imaging Receiver (SPIRE) by the HerMES survey \citep{Oliver2012} at $250 \,\rm \mu m$, $350\,\rm \mu m$, and $500\,\rm \mu m$; a Submillimetre Common-User Bolometer Array 2 (SCUBA-2) observation taken by the S2COSMOS survey at $850\,\rm \mu m$ \citep{Simpson2019}; and  Atacama Submillimeter Telescope Experiment (ASTE) Astronomical Thermal Emission Camera (AzTEC) 1.1mm images taken from \citet{Aretxaga2011}. We highlight that all of these surveys fully cover the JELS field.
We also utilized ALMA continuum maps from the A3COSMOS project \citep{Liu2019}, specifically release version 20250312, which includes all data from the ALMA archive released before the 12th of March 2025. The data consist of (dirty-beam) continuum images and primary-beam images for 4287 pointings in the COSMOS field, all obtained from programs with different frequencies, depths, beam sizes, and pixel scales. A result of this varied data is that the ALMA analysis requires careful homogenization.
There are a couple of sources within both samples with individual detections across the different ALMA bands, typically at the $\sim3\sigma$ level. The strongest detection is a Paschen-$\beta$ emitter with a $5.8\sigma$ detection and a flux of $0.84\,\rm mJy$ in ALMA Band 7. 

\section{Methods}\label{sec:methods}
\subsection{SED fitting}
To characterize the two Paschen-line emitter samples, we estimated photometric redshifts and fundamental properties such as SFR, stellar mass, and extinction by performing SED fitting.
We performed the SED fitting using two codes. 
The first code, EAZY-Py \citep{Brammer2008}, is used to derive photometric redshift (photo-z) estimates and the physical parameters of individual galaxies such as stellar mass, SFR, and $A_{\rm V}$ for all sources in our full multi-wavelength catalogs \citep[for details see][]{Pirie2024}. 
We use the default Flexible Stellar Population Synthesis \citep[FSPS;][]{Conroy2009, Conroy2010} template set and perform photometric zero-point offset corrections utilizing the available spectroscopic redshifts. We also add a $5\%$ systematic uncertainty in quadrature during the template fitting analysis. In the subsequent analysis, we use the EAZY values when analyzing individual galaxies.
The second code is STARDUST \citep{Kokorev2021}, which enables the extraction of galaxy properties by fitting their photometry using a set of linearly combined templates. These templates characterize the UV+Optical emission from dust-unobscured stellar light, AGN-heated dust in the mid-IR, and IR dust reprocessed stellar light in the near-IR and FIR. A feature of STARDUST is that it does not enforce energy balance, so the dust-obscured and unobscured components are not coupled in the fit. As we had already removed potential sources with strong AGN contributions from our sample, we have chosen to exclude the AGN model from our fits.
We use STARDUST to obtain the weighted median properties such as stellar mass and SFR of the two Paschen emitter samples by fitting the weighted median photometry of said samples. We also use STARDUST to constrain the weighted median dust properties such as dust mass, gas mass, and SFR for our stacked samples. The reason that we use the weighted median photometry as input for the SED fitting is to incorporate the uncertainty in the photometry when estimating the stacked physical parameters.

\subsection{Emission line star formation rates}
We estimate extinction-corrected line SFRs for the two Paschen-line samples.
We use the relations from \citeauthor{Reddy2023} (\citeyear{Reddy2023}; see also \citealp{Cleri2022} for a similar approach), assuming a subsolar stellar metallicity of $Z_{\rm \star} = 0.001$.
We apply the same F444W aperture correction to the line fluxes used for the continuum photometry.
For the extinction correction $10^{0.4\times A_{\rm Line}}$, we utilize the $A_V$ values from EAZY (Pa-$\alpha$ median $ A_{\rm V,med}=0.32^{+0.29}_{-0.18}$, Pa-$\beta$ median $ A_{\rm V,med}=0.24^{+0.52}_{-0.17}$) to estimate the nebular extinction for a given emission line. Following \cite{Reddy2020}, we adopt a \cite{Calzetti2000} attenuation law for the stellar continuum ($ R_{\rm V,\star}=4.05$) and a \cite{Cardelli1989} law for the nebular emission ($R_{\rm V,neb}=3.10$) together with the canonical \cite{Calzetti2000} relation $E(B-V)_{\rm\star}=0.44E(B-V)_{\rm neb}$ which aligns with other studies over a wide redshift range \citep[see Table 1 in][]{Shivaei2020} to obtain the relation $A_{\rm V,neb}=1.74\times A_{\rm V,\star}$, where $A_{\rm V,neb}$ is the nebular $A_{\rm V}$ and $A_{\rm V,\star}$ is the stellar $A_{\rm V}$. We then use the \cite{Cardelli1989} curve to obtain the ratio between $A_{\rm Line}/A_{\rm V,neb}$ to estimate $A_{\rm Line}$, which is nebular line extinction. The full SFR calculation is then $SFR_{\rm Line}=F_{\rm Line}\times10^{0.4\times A_{\rm Line}}$ where $F_{\rm Line}$ is the line flux.

\subsection{Weighted median stacking}\label{sec:stack}
To analyze the FIR properties of the narrowband-excess galaxies, we performed a stacking analysis \citep[see][]{Kurczynski2010, Bethermin2012a, Heinis2013, Ibar2013, Guijarro2022, Wang2022, Blanquez2023}. We first extracted cutout images centered on each source.
We then performed weighted-median stacking, which for a given stack of pixels consists of sorting their values and weights by their pixel intensity, then computing the cumulative weights and finding the point where half of the total cumulative weight value is reached, and lastly finding the index where the cumulative weight exceeds half the total weight (which is the median). This is more robust to outliers and non-Gaussian noise than a simple mean. This approach is especially relevant for the ALMA data we stack, as it has different root-mean-square (RMS) and beam sizes.
For Spitzer and Herschel images, the weights are derived from the noise maps, such that each pixel is assigned a weight $w_{\rm i}=1/\sigma_{\rm i}^2$ where $\sigma_{\rm i}$ is the noise at a given pixel. 
For ALMA data the primary beam image has to be taken into account when performing the stacking, so the pixel weights will instead be $w_{\rm i}=pb_{i}^2/\sigma_{\rm i}^2$ where $pb_{\rm i}$ is a given pixel value in the primary beam image and $\sigma_{\rm i}$ is the standard deviation of the $3\sigma$ clipped cutout pixels from the pre-primary beam corrected images.

We found that, especially for the MIPS, PACS, and SPIRE cutouts, the distribution of pixel intensities was significantly offset from zero 
due to the presence of the background. We would expect the pixels to have positive and negative intensities in roughly equal measure, with potential sources skewing the number of positive intensities. To account for this, we perform background subtraction on the images by subtracting the $3-\sigma$ clipped median value of the individual cutouts before stacking them.

At FIR wavelengths, the coarse resolution allows neighboring sources to blend within the beam. We therefore merged sources that overlap within the instrument Full Width at Half Maximum (FWHM) and stacked them using a single cutout centered on their mean position.

For the JWST and HST data, we estimated the uncertainty on the weighted-median flux with a bootstrap of 40,000 draws using the corrected 600 mas aperture fluxes. We also folded in a $5\%$ calibration error in quadrature.

For the MIPS, SPIRE, and AzTEC data, to extract fluxes, we performed PSF fitting using a 2D circular Gaussian from the Python package photutils \citep{Bradley2024}. We locked the position of the fit to the center of the stacked image and locked the FWHM of the PSF to the instrument band FWHM plus the size of one pixel. 
We took the peak value of the PSF model fit as the signal value. 
This approach is valid because we assumed the sources to be point-like. 
Following the results of \cite{Downes2012}, we added a flat 10\% increase to the fluxes and errors of the stack of the cutouts from the ASTE AzTEC map.

We estimated stacked flux densities and their associated uncertainties following standard methodology used in Herschel/SPIRE stacking analyses \citep{Bethermin2012a, Viero2013, Schreiber2015}. For each source bin, we applied the full stacking pipeline including weighting, background subtraction, and PSF fitting at the catalog positions of the sources in that bin, adopting the weighted median as our primary stacking statistic, since median stacking is less sensitive to outliers and to the contribution of rare bright sources than mean stacking \citep{Bourne2012}. For comparison, we also performed a weighted mean stack as a consistency check.
To assess the bias introduced by source clustering and other systematics inherent to the stacking estimator, we performed 1000 random position realizations using the identical pipeline (weighting, background subtraction, and PSF fitting) and the same number of stacked positions as in the real source catalog, drawn randomly within the survey footprint. 
The resulting distribution of recovered fluxes characterizes the null hypothesis in which no true astrophysical signal is associated with the stacked positions, naturally incorporating instrumental noise and confusion noise \citep{Kurczynski2010}. 
Since the random-position realizations were performed directly on the real SPIRE signal maps, confusion noise is naturally captured in the empirical scatter of the null distribution without requiring an explicit confusion noise term to be added in quadrature to the instrumental noise weights, as is necessary when propagating per-source noise estimates analytically \citep[e.g., ][]{Bourne2012, Smith2017}. 
Our approach is consistent with the empirical treatment of \cite{Bethermin2012a} and \cite{Viero2013}, in which bootstrap resampling of the real source catalog implicitly incorporates the confusion noise contribution to the stacking uncertainty without requiring it to be specified separately.
We took the median of this null distribution as our estimate of the residual bias, $S_{\rm bias}$, and corrected the real stacked flux as $S_{\rm corrected} = S_{\rm stack} - S_{\rm bias}$.
Uncertainties on the corrected stacked flux densities were estimated using a bootstrap resampling technique applied to the real source catalog, rather than the random-position distribution, since this captures the additional scatter associated with the specific sample being stacked \citep[e.g., sample size, source-to-source variation in flux, and any clustering enhancement specific to the real positions;][]{Bethermin2012a, Viero2013}. 
For each bin containing N sources, we drew N sources with replacement from the catalog and repeated the full stacking procedure (median or mean, as appropriate) on the resampled list, generating an empirical bootstrap distribution of 1000 stacked flux densities. We took the median of the bootstrap distribution to be $S_{\rm stack}$ and the $1\sigma$ uncertainty to be the standard deviation of this distribution.
For bins in which no statistically significant signal is detected, we report $3\sigma$ upper limits as the offset between the median of the random position distribution and the 99.865th percentile of the bootstrapped flux, the Gaussian-equivalent $3\sigma$ level: $S_{3\sigma} = P_{99.865}(S_{\rm corrected})$. 
This formulation avoids assuming Gaussianity for the underlying flux distribution, which may break down in the confusion-limited SPIRE regime \citep{Glenn2010}.

For the Herschel-PACS images, we
extracted fluxes using aperture photometry with a radius of 6.7 and 11.0 arcsec (their PSF FWHM) at 100 and 160 $\mu$m, respectively. We performed the bias estimate and bootstrapping using apertures of the same size in the image.
Aperture photometry is preferred in PACS mainly due to the uncertainties on the peak of the PSF introduced by the telemetry of the Herschel telescope and by the asymmetry seen along the scan directions.

For the ALMA data, we grouped images by band and resampled them to a common pixel scale. We then convolved the images to the smallest common beam using the Python package radio beam.\footnote{\url{https://github.com/radio-astro-tools/radio-beam}}
This was done so we have one data point for each ALMA band for the SED fitting, and so we are not mixing different beam sizes and pixel scales. Assuming a modified black-body, we also applied a K-correction to all images so that they are at a common frequency by assuming an emissivity index $\beta_{\rm IR}=1.8$ and a reference frequency that is the geometric mean of all the individual pointing frequencies.
Finally, we stack each group at its geometric mean frequency. 
Because some cutouts are missing, a simple stack could bias the result. We therefore applied multiple imputation to reconstruct incomplete datasets, assuming that the data are missing at random. This is combined with a cluster bootstrap to account for measurement uncertainties and sample variance.

Following on from similar work done by the ALPINE collaboration \citep{Khusanova2021}, we seek to produce a robust stacked image from a heterogeneous set of cutouts. However, data are missing for different subsets of galaxies, particularly in the ALMA coverage. In those cases, to fully propagate both measurement and sample selection uncertainties, we combined multiple imputations for missing sources and a cluster bootstrap over the sources. We began by grouping all available primary‐beam–corrected cutouts and their per‐pixel weights by source, as each source has one or more associated images. Because our true sample size \(N_{\rm total}\) exceeds the number of observed sources \(S_{\rm obs}\), we first performed multiple imputation. For each of \(M\) imputations we “hot‐deck” the \(N_{\rm miss}=N_{\rm total}-S_{\rm obs}\) missing sources by sampling (with replacement) from the observed source IDs, thereby generating a completed source list of length \(N_{\rm total}\) \citep{Rubin1987}. On the other hand, the cluster bootstrap runs over all galaxies with multiple images. This works under the assumption that the sources for which we have data are drawn randomly from the parent sample.

Within each imputed dataset, we applied a cluster bootstrap by drawing \(S_{\rm obs}\) source IDs with replacement from that completed list and including all images belonging to each drawn source in the bootstrap replicate.  For each of the resulting \(B\) replicates, we collapsed the selected images into a single stack \(Q_{\beta}(x,y)\) via a pixel‐wise weighted median.
After accumulating \(R=M\times B\) bootstrap‐imputed stacks, our final stacked image is the pixel‐wise median
\[
\rm Q_{\rm bar}(x,y)\;=\;MED(Q_{\beta}(x,y)),
\]
which automatically incorporates both the missing data uncertainty from imputation and the sampling variance from the cluster bootstrap.  
We used $M=20$ and $B=2000$ as the initial values. To minimize the Monte Carlo error to be less than $5\%$, we increased the number of imputations $M$ iteratively. 

To derive source fluxes for the ALMA stacks, we performed PSF fitting with a 2D Gaussian locked to the center of the stacks using the beam major-, minor-axis, and angle.    
Each of $R=M\times B$ images is measured to create a distribution of fluxes. In cases where the distribution's median is below a $3\sigma$ detection threshold, we report the 99.865th percentile as the $3\sigma$ upper limit for the ALMA stacks to account for non-Gaussian distributions. As a check, we also performed a simple weighted median stack on the ALMA images without any imputation or bootstrapping and inspected the stacked images for any $\geq 3\sigma$ peak detection. We also performed a weighted mean stack to see if there was any significant difference between the two approaches.
We highlight the FIR and ALMA fluxes in Tables \ref{tab:FIRflux} and \ref{tab:ALMAflux}. For comparison, we also have the weighted mean fluxes in Tables \ref{tab:meanFIRflux} and \ref{tab:meanALMAflux}.

There is a subset of JWST and Hubble data that has missing data. We employed the same hybrid multiple-imputation plus bootstrap approach as with the ALMA data on the corrected 600 mas aperture fluxes. 

\section{Results and analysis}\label{sec:Results}
\subsection{EAZY SED fitting and fundamental galaxy properties}
Utilizing the JWST/NIRCam and HST data, we performed SED fitting using EAZY to obtain photometric redshifts and fundamental properties such as stellar masses and SFR.
When spectroscopic redshifts were available (26/77 for Pa-$\alpha$ and 9/31 for Pa-$\beta$), the sources were fixed to those redshifts in the fit.
We visually inspected the SED fits of all Paschen line emitters to check if any particular fits failed or misidentified a galaxy.
We found that all fits broadly agree with the photometry and yield well-constrained redshift probability distributions with median reduced $\chi^2$ of $1.5^{+1.9}_{-1.0}$ for Pa-$\alpha$ and $1.7^{+1.7}_{-0.6}$ for Pa-$\beta$.
Our Paschen line emitter samples have the following stellar mass, SFR$\rm_{100\,Myr}$, and $A_{\rm V,\, Line}$ ranges:
\begin{itemize}
    \item[-] $7.36\leq \log(M_{\star}/M_{\odot})\leq 10.28$,
    \item[-]    $0.08<{\rm SFR}_{\rm 100\,Myr}/M_{\odot}\, {\rm yr}^{-1}<26.62$,
    \item[-]    $0.07<A_{\rm V}<2.65$
    \item[-]    $0.02<A_{\rm Pa-\alpha}<0.68$ for Pa-$\alpha$ and
    \item[-]    $8.19\leq \log(M_{\star}/M_{\odot})\leq 10.30$,
    \item[-]    $0.35<{\rm SFR}_{\rm 100\,Myr}/M_{\odot}\, {\rm yr}^{-1}<16.10$,
    \item[-]    $0.06<A_{\rm V}<1.90$
    \item[-]    $0.03<A_{\rm Pa-\beta}<0.89$ for Pa-$\beta$. 
\end{itemize}

\begin{figure*}
    \sidecaption
    \includegraphics[width=6cm]{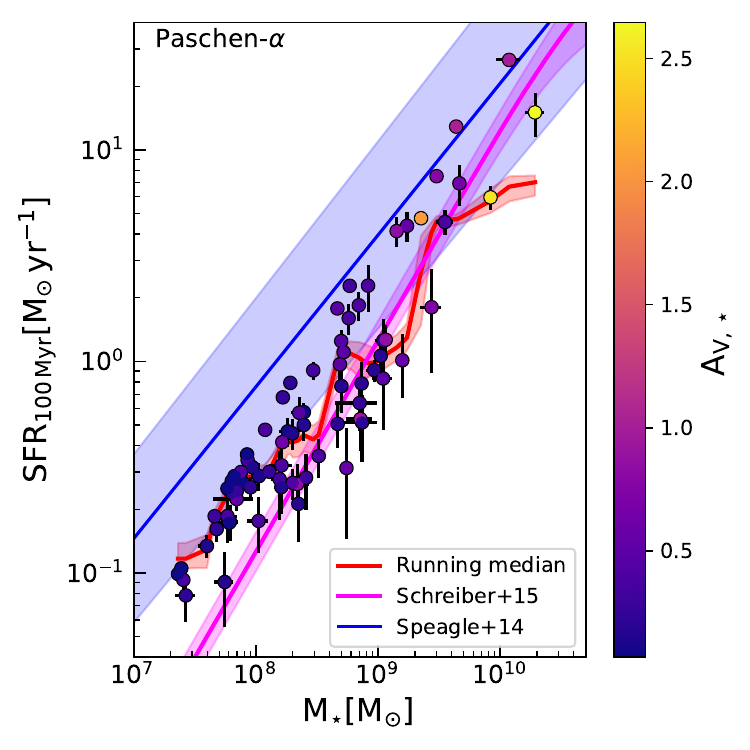}
    \includegraphics[width=6cm]{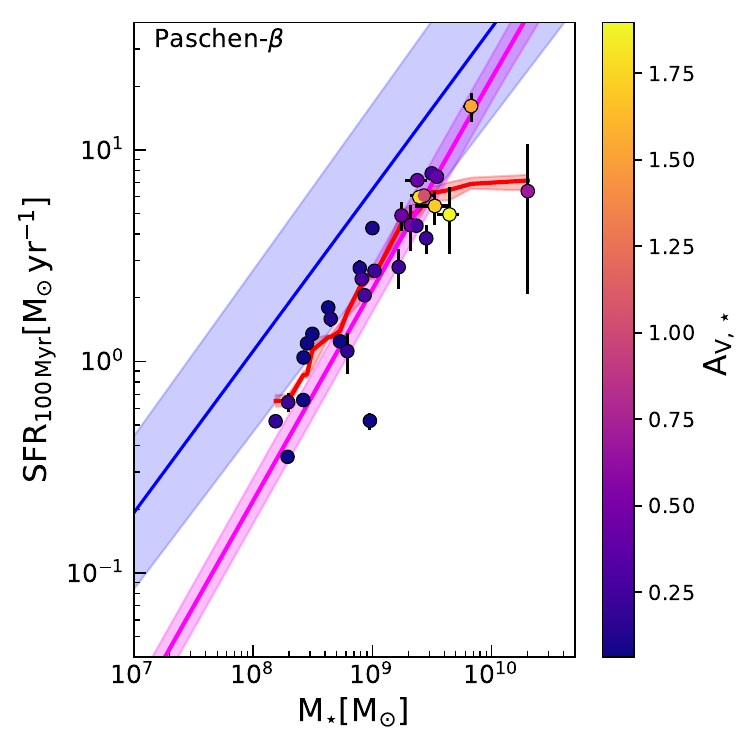}
    \caption{SFR vs. stellar mass for both Paschen line-emitter samples -- Pa-$\alpha$ (\textit{left}) and Pa-$\beta$ (\textit{right}) -- compared to literature galaxy main sequences. The solid blue and magenta lines indicate the star-forming main sequence at $z=1.5$ (left) and $z=2.65$ (right), with the shaded area being the 1$\sigma$ scatter \citep{Speagle2014, Schreiber2015}. 
    The Paschen line emitters are shown as colored circles, with the color being determined by their stellar $A_{\rm V}$ from EAZY SED fitting. A running median with an accompanying 1$\sigma$ error bar is shown in red.}
    \label{fig:MS}
\end{figure*}

We compared the two galaxy samples with the galaxy main sequence from \cite{Speagle2014} and \cite{Schreiber2015} at their respective redshifts in Fig. \ref{fig:MS}. We note that for \cite{Speagle2014} is the broad meta-fit over $10^{9.7}-10^{11}\,\rm M_{\odot}$ while \cite{Schreiber2015} is reliable over roughly $10^{9} - 10^{11}\,\rm M_{\odot}$ at $z\sim1.5$ and $10^{9.8}-10^{11}\,\rm M_{\odot}$ at $z\sim2.65$. Similar mass ranges are found in the literature \citep[see ][]{Whitaker2014, Pearson2018}.  
This means that for our lower-mass galaxies, the main sequences are extrapolations.
To identify trends in the data, we used a running median, taking the $k$-nearest neighbors. To capture both local and overall trends within the data, we use a value of $k\approx\sqrt{N}$, with N being the number of points in the sample. The error was determined from bootstrapping.
We find that both Paschen line-emitter samples scatter around the main sequence of \cite{Schreiber2015}. 
The Pa-$\alpha$ emitters appear to be slightly elevated from the \cite{Schreiber2015} main sequence at low masses and therefore closer to the \cite{Speagle2014} main sequence, while the most massive galaxy in the Pa-$\beta$ emitters sample is found below the main sequence. We discuss this massive Pa-$\beta$ emitter more in Sect. \ref{sec:Discussion}.
Generally, our selection appears well suited to target $\lesssim10^{9}\,\rm M_{\odot}$ galaxies on and above the galaxy main sequence.

To compare the SFR from SED fitting with the SFR estimated from the line emission, we needed the line fluxes. We find the fluxes to be in the range $\rm (2.61-636.71)\times10^{-19}\, erg\, s^{-1}\, cm^{-2}$ for Pa-$\alpha$ and $\rm (2.92-157.27)\times10^{-19}\, erg\, s^{-1}\, cm^{-2}$ for Pa-$\beta$ \citep[fluxes were estimated following eq. 3 and 6 in][]{Pirie2024}. After extinction correction, these correspond to line SFRs of $\rm 0.08-22.01\, M_{\odot}\, yr^{-1}$ and $\rm 0.68-47.86\, M_{\odot}\, yr^{-1}$, respectively.  
\begin{figure*}
    \sidecaption
    \includegraphics[width=6cm]{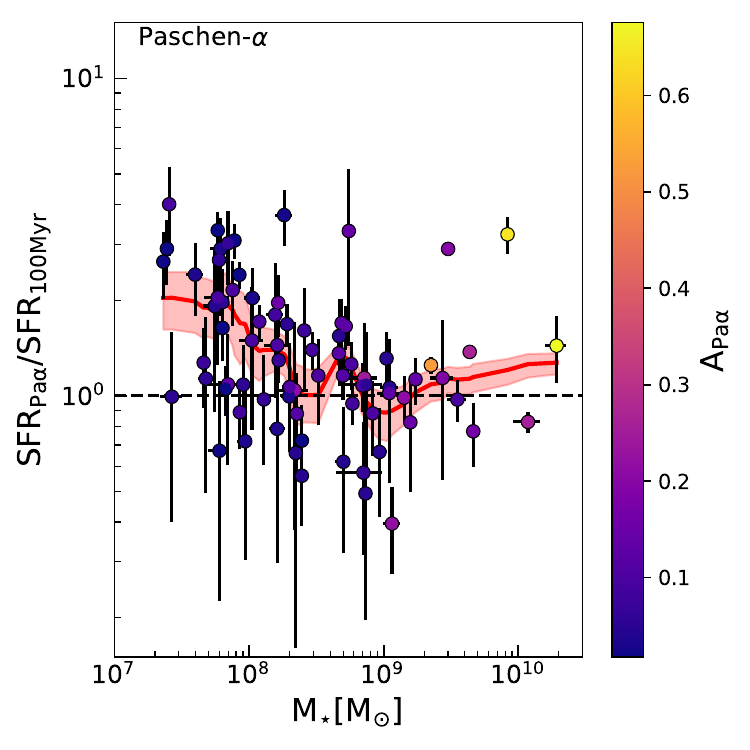}
    \includegraphics[width=6cm]{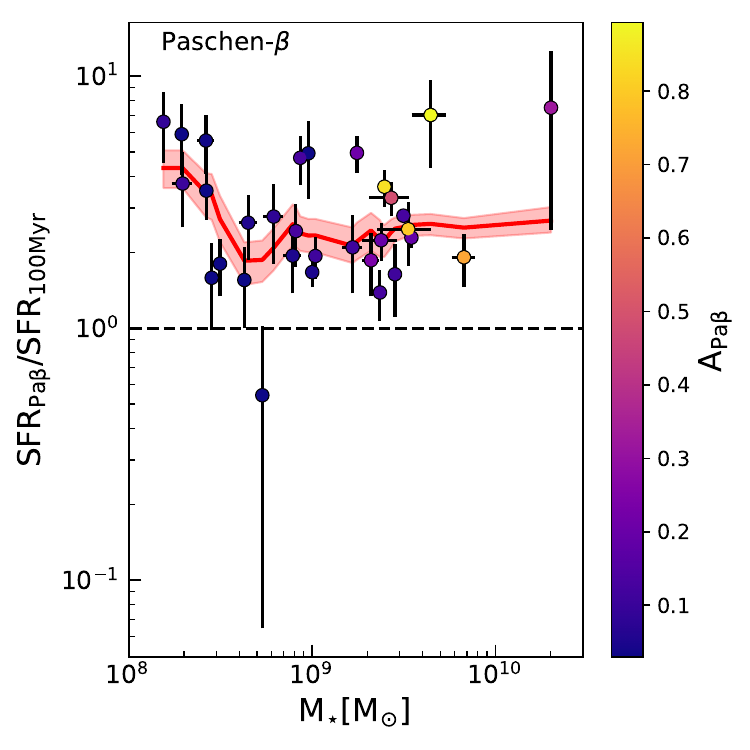}
    \caption{Ratio between the 100 Myr average from EAZY SED fitting and the Paschen line-SFR as a function of stellar mass for the  Pa-$\alpha$  (\textit{left}) and Pa-$\beta$  (\textit{right}) samples. Each point is colored according to its $\rm A_{\rm Line}$ value. We estimated $A_{\rm Line}$ by converting the $A_{\rm V,\star}$ value from the SED fitting using a \cite{Cardelli1989} curve. 
    The striped line indicates the 1:1 ratio. A running median with an accompanying 1$\sigma$ error bar is shown in red.}
    \label{fig:SBurst}
\end{figure*}
Taking the ratio between the 100 Myr average SFR from the SED fitting and the line-SFR allows for the burstiness (i.e., ${\rm SFR}_{\rm Line}/{\rm SFR}_{100\,\rm Myr}$, the ratio of short- to long-term star formation) of the galaxies to be assessed. The 100 Myr-averaged SFR is derived from the UV-to-near-IR EAZY SED fit to the broadband photometry; the FIR/submillimeter data are not included in this estimate and are treated separately in the stacking analysis.
Figure \ref{fig:SBurst} highlights the burstiness of the two samples as a function of stellar mass. Similar to the H$\alpha$ narrowband-selected sample at $z=6.1$ from \cite{Pirie2024}, essentially all our galaxies are found above or within the 1:1 line at the $1\sigma$ level. This suggests they are likely to be undergoing starbursts. Ten Pa-$\alpha$ emitters and one Pa-$\beta$ emitter are below the 1:1 line, even considering the $\rm 1\sigma$ uncertainty. 
We investigated whether there was anything else that set these emitters below the 1:1 line apart from the others. Inspecting the image cutouts, the sizes of these sources are comparable to the rest of the sample. A majority of the narrowband excess regions in these subsamples are compact, but this is not unique to these sources when compared to the full samples. We also inspected the EAZY SED fits, but found them to be similar to the other galaxies within the samples. 
The burstiness decreases slightly with stellar mass below $\sim10^9\,\rm M_{\odot}$ and then flattens at higher masses
for both samples.
To quantify the scale and significance of the trend, we calculated the Theil–Sen slope and Kendall’s $\tau$ with the log-log scaled data utilizing parametric Monte Carlo sampling. We used 2000 draws and assumed Gaussian errors, accounting for both errors and 100 permutations per draw to estimate the p value.
The slope is $-0.151^{+0.027}_{-0.027}$ (p value of 0.002), 
and Kendall’s $\tau$ is $-0.244^{+0.043}_{-0.042}$ (p value of 0.002) for Pa-$\alpha$,
while $-0.043^{+0.054}_{-0.055}$ (p value of 0.615), $-0.054^{+0.060}_{-0.065}$ (p value of 0.662) for Pa-$\beta$. 
For Pa-$\alpha$, the numbers indicate an overall statistically significant, monotonic downward trend. For Pa-$\beta$, the trend is not statistically significant; however, the confidence intervals include both small negative and small positive values, so the data are inconclusive with respect to very small effects. 

A downward trend could be expected as higher-mass galaxies have deeper gravitational potentials, more stable and ordered gas reservoirs, and (on average) steadier gas supply channels \citep{Dekel2006}. 
This results in stellar feedback (e.g., supernovae, winds) being less able to expel or strongly perturb the cold gas, and they suppress short, large amplitude swings in the SFR \citep{Faucher2018}.
In contrast, lower-mass galaxies have shallower potentials, leading to feedback having a greater ability to easily eject or redistribute gas \citep{Muratov2015} and producing repeated, strong bursts and lulls \citep{Weisz2012, El-Badry2016, Sparre2017}.
This picture is complicated by the fact that feedback forces scale approximately with the radius r as $\sim1/r^2$ while gravity scales roughly as $\sim1/r$, so whether feedback overcomes gravity depends on radius, gas distribution, and where star formation is concentrated — in some regions (e.g., compact centers or during mergers), feedback can still dominate and drive strong outflows. 
At lower stellar masses (and thus lower gas-phase metallicities), SNe are expected to be suppressed because the threshold for direct collapse shifts \citep[as argued by][]{Jecmen2023}, and radiation-pressure coupling is also less efficient at low metallicity.
These effects reduce the momentum and energy injection available to eject gas. Conversely, lower-metallicity stellar populations are harder ionizers (the interior iron bump drops), which can lead to bulk ionization and heating that suppresses star formation via over-pressurization rather than by ejecting gas \citep[see the discussion in Sect. 7 of][]{Flury2025}.
Therefore, the net outcome in low-mass, low-metallicity systems reflects a competition between shallower gravity (favoring ejection) and reduced and delayed mechanical coupling or increased radiation-driven heating (limiting it).
Curiously, for both line-emitter samples, some of the most massive galaxies appear clearly above the trend line. These outliers are discussed further in Sect. \ref{sec:Discussion}.

\begin{figure*}
    \sidecaption
    \includegraphics[width=6cm]{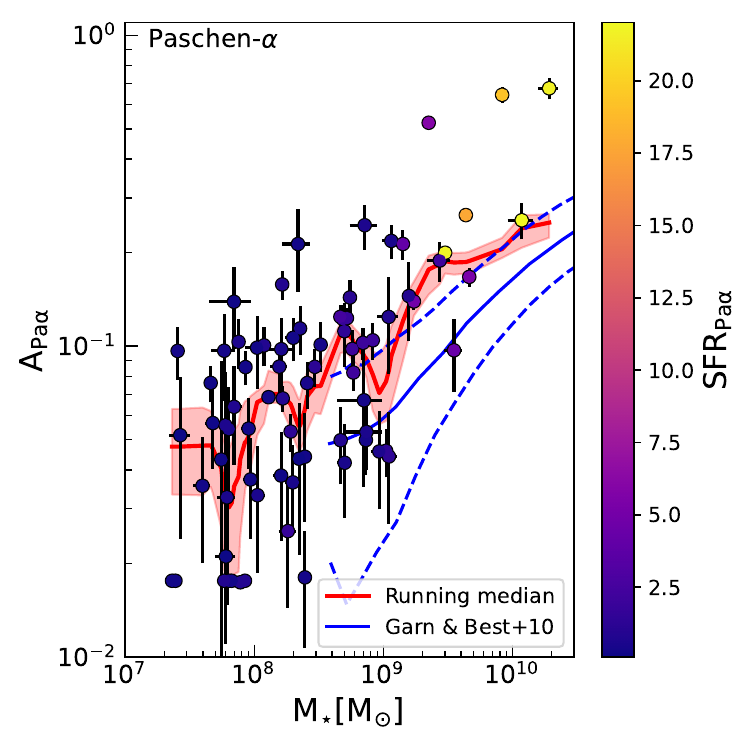}
    \includegraphics[width=6cm]{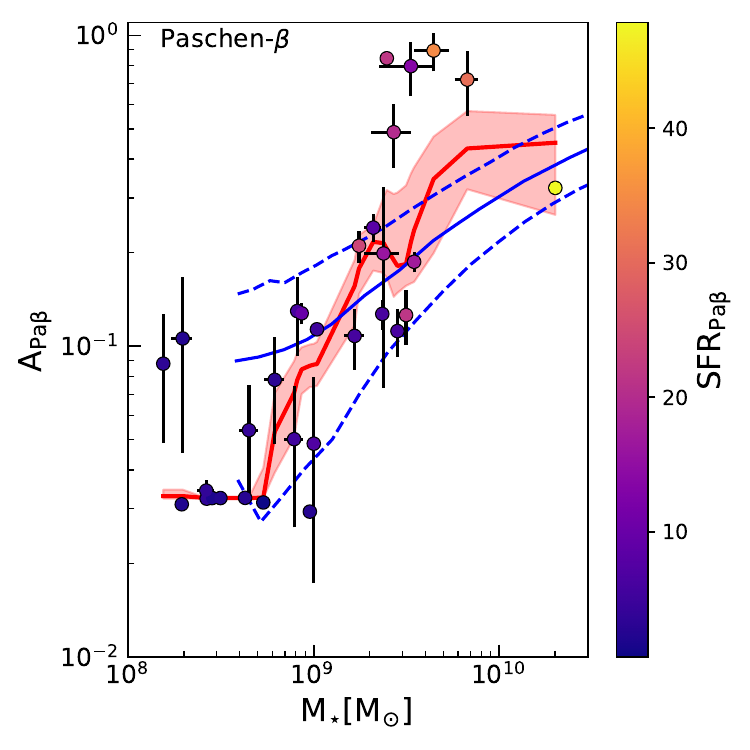}
    \caption{Relationships between $A_{\rm Line}$ and stellar mass for the Pa-$\alpha$  (\textit{left}) and Pa-$\beta$  (\textit{right}) samples. The $A_{\rm Line}$ was estimated by converting the $A_{\rm V,\star}$ value from SED fitting using a \cite{Cardelli1989} curve. 
    The points are colored by their line SFR. A running median with an accompanying 1$\sigma$ error bar is shown in red. For comparison, the median and standard deviation for the binned SDSS H$\alpha$ emitters of \cite{Garn2010b} are shown in blue. The \cite{Garn2010b} $A_{\rm H\alpha}$ values have been converted to Paschen line values by taking the ratio of the two $A_{\rm Line}$ values from the \cite{Cardelli1989} curve. A running median with an accompanying 1$\sigma$ error bar is shown in red.}
    \label{fig:ALine}
\end{figure*}

While there does not appear to be statistically strong evidence of a correlation between the burstiness and $A_{\rm Line}$ (see values in Table \ref{tab:fits}), there does appear to be a relation to the stellar mass, with higher-mass galaxies being more obscured. To test this, we plotted the $A_{\rm Line}$ as a function of stellar mass for the Paschen lines in Fig. \ref{fig:ALine}. For comparison, we included the median and standard deviation for the binned Sloan Digital Sky Survey (SDSS) H$\alpha$ emitters of \cite{Garn2010b}. The \cite{Garn2010b} $A_{\rm H\alpha}$ values have been converted to Paschen line values by taking the ratio of the two $A_{\rm Line}$ values from the \cite{Cardelli1989} curve. The curve is cut off at $\sim10^{8.3}\,\rm M_{\odot}$ as that is the mass limit for the \cite{Garn2010b} study.
We also present the relations between $A_{\rm Line}$ and $\rm SFR_{Line}$, $\rm SFR_{100\, Myr}$ in Figs. \ref{fig:ALine2} and \ref{fig:ALine3}. We show the slopes and Kendall's $\tau$ values in Table \ref{tab:fits}. All show statistically significant upward monotonic trends.
We see clear positive correlations for all cases and samples. Comparing with the \cite{Garn2010b} curve, we see general agreement with a monotonic increase with increasing mass. The Pa-$\alpha$ sample appears elevated compared to the SDSS galaxies, especially at the high mass end. For the Pa-$\beta$, there is better overall agreement, though there is still a group of five galaxies at the highest masses that appear elevated compared to the SDSS median.
These results of having positive correlations with $A_{\rm Line}$ are consistent with what has been found in the literature, such as \cite{Garn2010a, Garn2010b, Ibar2013, Kashino2013}, which showed similar relations for local H$\alpha$ emitters to $z\sim1.6$. Similar results have been shown for the UV attenuation \citep[see][]{McLure2018}.
These literature relations are found to, on average, describe the
properties of galaxies with different morphologies, bulge-to-disc ratios, galaxy environments, and both merging and non-merging galaxies.
\begin{table*}
\caption{Theil--Sen slopes and Kendall's $\tau$ with accompanying p values in parentheses for different relations in the Pa-$\alpha$ and Pa-$\beta$ samples.}
\label{tab:fits}
\centering
\setlength{\tabcolsep}{5pt}
\renewcommand{\arraystretch}{1.4}
\begin{tabular}{lcccc}
\hline\hline
Relation &
Pa-$\alpha$ slope &
Pa-$\alpha$ $\tau$ &
Pa-$\beta$ slope &
Pa-$\beta$ $\tau$ \\
\hline
$\frac{SFR_{\rm Line}}{SFR_{\rm 100\,Myr}}$
vs. $\mathrm{M}_\star$
&
$-0.151^{+0.027}_{-0.027}$ (0.002)
&
$-0.244^{+0.043}_{-0.042}$ (0.002)
&
$-0.043^{+0.054}_{-0.055}$ (0.615)
&
$-0.054^{+0.060}_{-0.065}$ (0.662)
\\

$\frac{SFR_{\rm Line}}{SFR_{\rm 100\,Myr}}$
vs. $A_{\rm Line}$
&
$-0.141^{+0.111}_{-0.122}$ (0.370)
&
$-0.068^{+0.048}_{-0.046}$ (0.383)
&
$0.210^{+0.216}_{-0.179}$ (0.385)
&
$0.103^{+0.068}_{-0.069}$ (0.427)
\\

$A_{\rm Line}$
vs. $\mathrm{M}_\star$
&
$0.344^{+0.024}_{-0.025}$ ($5\times10^{-6}$)
&
$0.420^{+0.029}_{-0.032}$ ($5\times10^{-6}$)
&
$0.706^{+0.048}_{-0.044}$ ($2.0\times10^{-5}$)
&
$0.547^{+0.047}_{-0.047}$ ($5\times10^{-6}$)
\\

$A_{\rm Line}$
vs. $SFR_{\rm Line}$
&
$0.340^{+0.038}_{-0.034}$ ($5.5\times10^{-5}$)
&
$0.339^{+0.032}_{-0.037}$ ($2.5\times10^{-5}$)
&
$0.864^{+0.071}_{-0.062}$ ($5\times10^{-6}$)
&
$0.624^{+0.052}_{-0.056}$ ($5\times10^{-6}$)
\\

$A_{\rm Line}$
vs. $SFR_{\rm 100\,Myr}$
&
$0.374^{+0.038}_{-0.035}$ ($1\times10^{-5}$)
&
$0.373^{+0.033}_{-0.033}$ ($<2.5\times10^{-6}$)
&
$0.830^{+0.064}_{-0.060}$ ($3.5\times10^{-5}$)
&
$0.555^{+0.052}_{-0.052}$ ($1.5\times10^{-5}$)
\\
\hline
\end{tabular}
\tablefoot{Estimates were derived using logarithmic quantities and Monte Carlo sampling to account for uncertainties in both variables. The smallest p value that can be resolved given the number of draws and permutations is $2.5\times10^{-6}$.}
\end{table*}

\begin{figure*}
    \sidecaption
    \includegraphics[trim={0 0 10cm 0},clip,width=12cm]{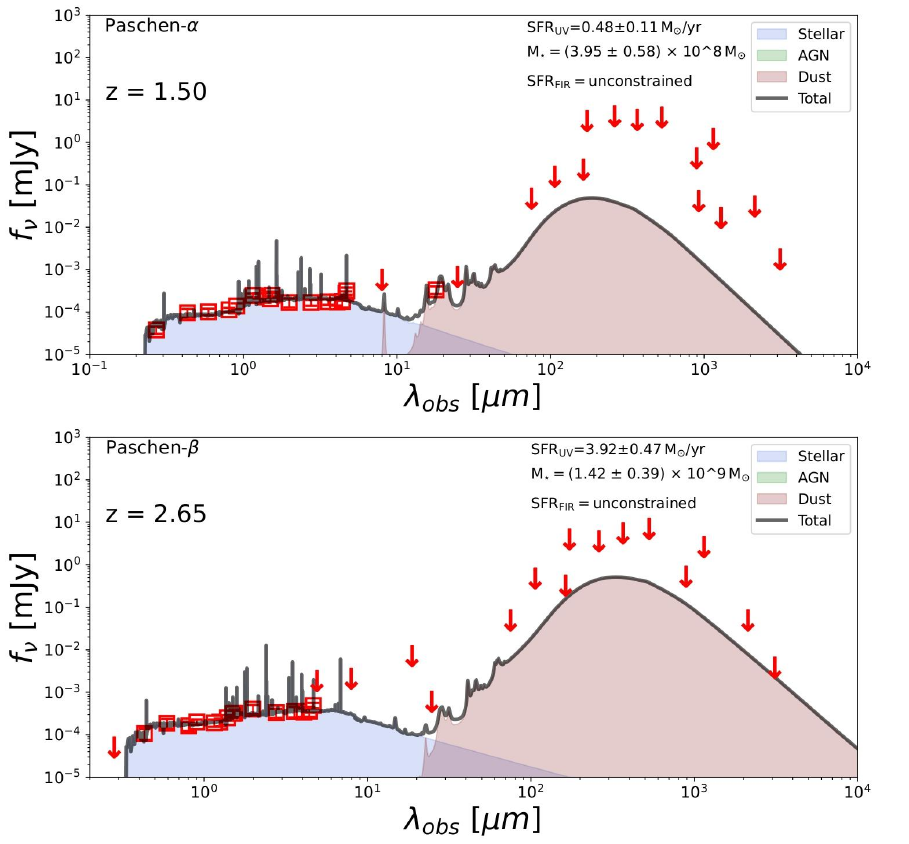}
    \caption{SED fits of the weighted median stacked photometry utilizing the STARDUST code \citep{Kokorev2021} for the two Paschen line-emitter samples: Pa-$\alpha$  (\textit{top}) and Pa-$\beta$  (\textit{bottom}). Photometric points are shown as red squares with error bars, and $3\sigma$ upper limits are shown as red arrows, with the top of the arrow being the limit. The different components of the SED are shown with different colors: blue for stellar, green for AGN, and red for dust. The total SED shown in black. The fits were performed without AGN contribution. Although we include the dust emission template to fully explain the MIRI emission, the FIR is essentially unconstrained.}
    \label{fig:STARDUST}
\end{figure*}
\subsection{The obscured star formation}
To estimate the IR SFR and gas mass, we need to use the weighted median stacked fluxes to perform SED fitting.
The resulting weighted median stacked fluxes are shown in Tables \ref{tab:JWSTflux}-\ref{tab:ALMAflux}.  
The STARDUST SED fits using the weighted median stacked photometry for the Pa-$\alpha$ and Pa-$\beta$ samples are shown in Fig. \ref{fig:STARDUST}. The redshift of the SED fits was set to the center of the redshift probed by the narrowband filters for both samples ($z=1.5$ for Pa-$\alpha$ and $z=2.65$ for Pa-$\beta$).
For the Pa-$\beta$ sample, we chose not to include the upper limits from ALMA Bands 6 and 7 because they are highly uncertain, as only three and five galaxies have data in those bands, respectively.
We find that for the Pa-$\alpha$ sample, none of the bands at wavelengths higher than MIRI F1800W have a $\geq3\sigma$ detection, while for the Pa-$\beta$ sample, it is for wavelengths longer than the narrowbands. 
While we included the dust emission template to fully 
explain the MIRI emission, the FIR is essentially unconstrained. 
Our deepest and most complete constraint in the FIR comes from ALMA Band 4, which includes all the data from the ALMA MORA project \citep{Casey2021}, covering both samples. For the Pa-$\beta$ sample, the weighted median and mean stacked images show a potential detection. The median image has a tentative $2.67\sigma$ peak $1.38''$ offset from the central position, but within the beam, while the mean image has a slightly higher $3.76\sigma$ detection $1.10''$ offset from the center (see Fig. \ref{fig:Band4stacks}). Given the offset from the center and the relatively low S/N of the peak, we used the bootstrap upper limits for the rest of the analysis. A deeper ALMA follow-up of the dust continuum in individual sources would be able to more securely constrain the dust properties.  

Because the SED fitting is unconstrained in the FIR, we estimated the $3\sigma$ upper limit of the IR SFR by assuming an optically thin modified blackbody (gray body) spectral shape and normalizing its amplitude to the ALMA Band 4 limit. While some of the other ALMA bands appear deeper, such as Band 3, the lack of data for every galaxy makes relying on the limits for analysis riskier, as we might be missing information about the sample variability. This is true even considering that the bootstrapping and imputation procedure tries to account for the effect of missing data. 
To estimate the dust temperature $T_{\rm dust}$, we used Eq. 5 from \cite{Magnelli2014}, which depends on the specific SFR (sSFR):
\begin{equation}
T_{\rm dust}=98\times(1+z)^{-0.065}+6.9\times \log({\rm sSFR}).
\end{equation}
We used the UV-optical SFR from the stellar part of the STARDUST fit to calculate the sSFR. Given the $\rm SFR_{UV}$ of $0.48\pm0.11\,\rm M_{\odot}\,yr^{-1}$ and $3.92\pm0.47\,\rm M_{\odot}\,yr^{-1}$ and stellar masses of $(3.95\pm0.58)\times10^{8}\,\rm M_{\odot}$ and $(1.42\pm0.39)\times10^{9}\,\rm M_{\odot}$ for the Pa-$\alpha$ and Pa-$\beta$ samples, respectively, we obtain dust temperatures of $\rm31\, K$ for both. These values are typical for star-forming galaxies \citep{Magnelli2014}.
The blackbody dust emissivity spectral index $\rm \beta$ typically varies between $1.5-2.0$ \citep{Kennicutt2011, Bethermin2015}, and we therefore assumed $\rm \beta=2.0$ as an upper limit. 
We estimated the rest-frame $8-1000\,\mu$m IR luminosity and convert to an IR SFR following the \cite{Kennicutt2012} conversion for a \cite{Chabrier2003} initial mass function: ${\rm SFR}_{\rm IR}=1.0\times10^{-10}\times L_{\rm IR}$. We find 3$\sigma$ upper limits for Pa-$\alpha$ sample of ${\rm SFR}_{\rm IR}\lesssim53\,\rm M_{\odot}\,yr^{-1}$ and for Pa-$\beta$ sample of ${\rm SFR}_{\rm IR}\lesssim44\,\rm M_{\odot}\,yr^{-1}$. When comparing to the UV-optical SFR from the STARDUST fit, the ${\rm SFR}_{\rm IR}$ limits are more than one dex higher. This means that while there is no evidence of strong dust emission from the stacked data,
there is potentially a significant amount of the total SFR to be missed by not accounting for the dust emission. In the future, deeper targeted ALMA observations could constrain this discrepancy between the IR and UV SFR better. 
Individual detections highlight this further, the two radio-detected galaxies (one Pa-$\alpha$ and one Pa-$\beta$ emitter) have ${\rm SFR}_{\rm IR}$ that are $\sim 3.5\times$ and $\sim64\times$ higher than ${\rm SFR}_{100\,\rm Myr}$ and $\sim2.8\times$ and $\sim8.5\times$ higher than ${\rm SFR}_{\rm Line}$, respectively.
The gas mass upper limits are similarly too weak to determine the typical gas reservoir of these galaxies and their gas depletion time ($M_{\rm gas}\rm\lesssim10^{11.2}\,M_\odot$ for Pa-$\alpha$ and $M_{\rm gas}\rm\lesssim10^{10.6}\,M_\odot$ for Pa-$\beta$, see Appendix \ref{sec:Gas mass} for details).

\section{Discussion}\label{sec:Discussion}
\subsection{Potential biasses}
To investigate potential sample selection biases,
we performed two-sample Kolmogorov–Smirnov (K-S) tests using the stellar masses and sSFRs, with the null hypothesis that the two independent Paschen samples are drawn from the same continuous distribution as the field under a chosen confidence level of 95\%. Our galaxies are contrasted with the galaxies detected in the JELS field that are not narrowband excess sources and are within the same redshift ranges that the narrowbands probe.
For Pa-$\alpha$, we find that this comparison results in stellar masses having a KS value of $0.30$ and a p value of $4.60\times10^{-3}$, and the sSFRs have a KS value of $0.66$ and a p value of $2.13\times10^{-13}$, meaning that in both cases we can reject the null hypothesis and state that the galaxies are drawn from separate parent samples. 
For Pa-$\beta$, we find that the stellar masses have a K-S value of $0.24$ and a p value of $0.27$, and the sSFRs have a K-S value of $0.46$ and a p value of $1.32\times10^{-3}$, meaning we cannot reject the null hypothesis when it comes to the stellar masses, but we can for the sSFR. So, while the Pa-$\beta$ does not appear to have a mass bias compared to the field, the Pa-$\alpha$ sample has an overabundance of low-mass galaxies. Both samples are biased toward higher sSFRs than the field, or put another way, they are biased toward being above or on the upper end of the main sequence, as seen in Fig. \ref{fig:MS}. 
The stellar mass cutoff for the $z\approx2.65$ galaxies in the JELS is $\sim2\times10^8\,\rm M_{\odot}$, which is about the point below which the Pa-$\alpha$ starts to be significantly above the main sequence. This could suggest that a similar population might be missing from the Pa-$\beta$ sample due to the current lack of sensitivity. 
It is somewhat unsurprising that the samples appear biased toward higher sSFR than the rest of the JELS field, as higher sSFR means more ongoing star formation per unit stellar mass, which boosts the emission-line flux relative to the stellar continuum, increasing the equivalent width and making the line emitter easier to detect.

We expect that it is generally harder to select narrowband excess sources as stellar mass increases. The stellar continuum is stronger in higher-mass galaxies due to older stellar populations in massive galaxies that add continuum light but do not add much line emission, and massive galaxies generally have lower sSFR. This results in lower equivalent widths for higher-mass galaxies.
As a consequence, we would expect that the percentage of narrowband excess detected compared to the rest of the JELS field decreases as a function of mass.
We tested this by binning the stellar masses of excess sources and the other galaxies in the JELS field and calculating the percentage of narrowband excess sources within each bin.
We found that Pa-$\alpha$ has $\approx65\%$ up until $\sim10^9\,\rm M_{\odot}$, after which there is a decreasing trend with stellar mass. 
For Pa-$\beta$ we see greater fluctuations in the percentage, between $30-80\%$ until $\sim10^{10}\,\rm M_{\odot}$.
Both samples tend to zero at the high mass end above $\sim10^{10}\,\rm M_{\odot}$ as there are no or very few excess sources at those masses (see Fig. \ref{fig:Cratios}). 

We also tested how low a burstiness we can probe in main-sequence galaxies as a function of stellar mass. We did this by taking the lowest line flux in our sample and converting it to a line SFR. We then used the \cite{Schreiber2015} main sequence to estimate the 100 Myr average SFR for a typical main sequence galaxy as a function of stellar mass. Only 5/77 Pa-$\alpha$ sources are within $1\sigma$ of the inferred detection boundary, and the rest are above it, indicating that the flux limit affects only a small fraction of the sample. While incompleteness may still be important at the lowest masses, it cannot alone explain the trend we observe.
For the Pa-$\beta$ sample, the line is below the 1:1, and none of the sources are within $1\sigma$ of the line, meaning the increased burstiness compared to the Pa-$\alpha$ cannot be explained by the survey flux limit. What is more likely the culprit in the difference between the two samples is the line luminosity to SFR calibration from \cite{Reddy2023}.

When comparing the two samples in Fig. \ref{fig:SBurst}, we see that the Pa-$\beta$ sample has an elevated short-term SFR compared to the Pa-$\alpha$ sample, being around two times higher. This is interesting because it aligns with the difference in the line flux to SFR calibration of \cite{Reddy2023} that we applied, where for ${\rm SFR_{Line}[M_\odot\, yr^{-1}] = C_{Line} \times L_{Line}[erg\, s^{-1}]}$ we have ${\rm C_{Pa\beta}/C_{Pa\alpha}=1.97}$. We note that in the literature, a solar metallicity value is often used for the calibration, which is a factor $\sim2\times$ higher \citep{Kennicutt1998, Kennicutt2012, Arteaga2022, Reddy2023}. Since we do not have access to the $A_{\rm Line}$ directly, we are unable to assess the degree to which our SFR estimates align with the \cite{Reddy2023} calibrations. In the future, spectroscopic follow-up of the line emitters with instruments such as NIRSpec on JWST will be able to inform us of the obscuration of the Paschen line emitters in detail. There has recently been some evidence in the literature that suggests the nebular $\rm R_{V}$ varies from the \cite{Cardelli1989} value, with a median $\rm R_{V}$ closer to $\approx6$ for $z>1.5$ galaxies \citep{Reddy2026}. This would in turn increase the nebular $\rm A_{Line}$, resulting in our line SFRs being higher when correcting for extinction.     
\subsection{Possible massive galaxy rejuvenation for Paschen line emitters}
When comparing Figs. \ref{fig:MS} and \ref{fig:SBurst}, we find that, interestingly, two of the Paschen-$\alpha$ and two of the Paschen-$\beta$ emitters that are the most massive also appear to be among the sources undergoing the strongest burst, with high Paschen line emission compared to the 100 Myr-averaged SFR. 
Most of these galaxies appear to scatter around the \cite{Schreiber2015} galaxy main sequence and are within $1\sigma$ of it. However, the most massive Paschen-$\beta$ emitters, with an sSFR of $\rm (3.18\pm2.14)\times10^{-10}\,yr^{-1}$ and a position $0.83+/-0.44$ dex below the \cite{Schreiber2015} main sequence, could potentially be quiescent depending on the classification \citep{Donnari2019, Morselli2019}.  
This suggests a rejuvenation in star formation after a quiescent period. As we would generally expect these high stellar mass galaxies to have less strong bursts than their lower-mass counterparts, this rejuvenation might come as a result of merger activity \citep{Pearson2019, Harrold2026}. The deeper gravitational potential of the high stellar mass galaxies attracts matter, which results in minor mergers that trigger star formation. Another culprit could be cold streams of gas that penetrate the hot halos of the massive galaxies and fuel star formation \citep{Dekel2009}. A deeper analysis of the morphology of these galaxies will be done in the future and will allow us to constrain the factors behind the rejuvenation of star formation in detail (Ossa-Fuentes et al. in prep).

\section{Conclusions}\label{sec:Conclusions}
Using the JELS F466N/F470N narrowband observations embedded within the PRIMER multiband imaging of COSMOS, we selected robust samples of Paschen-$\alpha$ (77 $z\sim1.50$ sources across the two filters) and Paschen-$\beta$ (31 $z\sim2.65$ sources) emitters. We adopted conservative color-excess, photometric redshift, and visual-inspection criteria to minimize contaminants for the purpose of stacking analysis. We assembled a multi-wavelength dataset covering JWST/NIRCam+MIRI, HST, Spitzer MIPS, Herschel PACS/SPIRE, SCUBA-2, AzTEC, and ALMA, and we adopted band-specific extraction procedures (PSF fitting, aperture photometry, background subtraction) tuned to each instrument’s characteristics. 
For ALMA, we grouped heterogeneous image products by band, homogenized pixel scales and beams, and implemented a robust image-plane weighted-median stacking procedure that combines multiple imputation for missing cutouts with a cluster bootstrap to propagate both measurement and sample-selection uncertainties. We report the following conclusions:

\begin{enumerate}
\item EAZY SED fits indicate our Paschen-emitter samples span a wide range of stellar masses ($\log(M_\star/M_\odot) \sim 7.4$--$10.3$) and modest SFRs ($\sim 0.1$--$50\,\rm M_\odot\,yr^{-1}$). The samples broadly scatter around the literature main sequence at their respective redshifts.
\item Comparing the short-timescale Paschen line SFRs to the 100 Myr-average SED SFRs revealed widespread burstiness, with most sources lying above the 1:1 line, and a weak negative trend of burstiness with stellar mass is apparent for the Paschen-$\alpha$ sample, which is consistent with theoretical expectations for stronger stochasticity in low-mass systems. In contrast, the Paschen-$\beta$ sample shows no statistically significant trend.
\item We find clear, positive correlations between nebular attenuation ($\rm A_{\rm Line}$) with stellar mass and SFR (line and SED). This outcome is qualitatively consistent with previous studies of emission-line selected samples and implies increasing dust obscuration in more massive and more actively star-forming systems.
\item Despite the comprehensive FIR/submillimeter stacking, we did not detect any stacked emission above $3\sigma$ at wavelengths longer than MIRI.  
Using the Band 4 constraints, we derived $3\sigma$ upper limits on $\rm SFR_{IR}$ of
$\lesssim 53\,M_\odot\,$yr$^{-1}$ (Pa-$\alpha$) and $\lesssim 44\,M_\odot\,$yr$^{-1}$ (Pa-$\beta$), more than one order of magnitude higher than the SFR from fitting the optical and near-IR photometry. Similarly, the gas masses are too uninformative to constrain the gas mass ratios or the depletion times. 
\item Both the Pa-$\alpha$ and Pa-$\beta$ samples are preferentially drawn from galaxies with elevated sSFRs, reflecting the inherent bias of emission-line selection toward systems with high equivalent widths. In addition, the Pa-$\alpha$ sample exhibits a significant bias toward lower stellar masses, whereas the Pa-$\beta$ sample does not. The fraction of narrowband excess sources declines with increasing stellar mass, which is consistent with the decreasing equivalent widths expected in more massive galaxies. Sensitivity limits may contribute to incompleteness at the lowest masses, particularly for Pa-$\alpha$, but they cannot fully account for the observed trends.
\item A subset of the highest-mass Paschen line emitters have relatively strong bursts, with the most massive Pa-$\beta$ emitter potentially being classified as quiescent. This combination is suggestive of recent rejuvenation of star formation (e.g., minor mergers or gas accretion). Morphological analysis and spectroscopic follow-up are required to discriminate between merger-driven and cold-accretion scenarios.
\end{enumerate}

In summary, narrowband Paschen line-emitter selection with JWST provides a complementary probe of short-timescale star formation across a wide mass range, but current FIR/submillimeter archival data place only upper limits on the typical dust-obscured SFR and gas masses of the samples. Deeper ALMA continuum measurements (preferentially in Bands 6 and 7) and spectroscopic metallicity diagnostics are the most direct paths to tightening constraints on obscured star formation, dust content, and gas reservoirs in these Paschen line-emitter samples and thus to test how much dusty star formation there is when the FIR data are taken into consideration.

\begin{acknowledgements}
We thank the A3COSMOS collaboration, specifically Sylvia Adscheid and Daizhong Liu, for initially providing us with access to the images from version 20220606 of A3COSMOS. 
We thank Juan Molina for the fruitful discussion we had on the removal of AGNs within the Paschen samples.
We thank Ian R. Smail for the helpful comments that they provided.
We also gratefully acknowledge financial support from ANID - MILENIO - NCN2024\_112 and funding by project ALMA-ANID N$^\circ$31230049. E.I. acknowledges funding from ANID FONDECYT Regular 1221846. L.O.\ acknowledges the National Doctoral Degree Scholarship given by the National Research and Development Agency of Chile (ANID, grant number 21220499).
RKC is grateful for support from the Leverhulme Trust via the Leverhulme Early Career Fellowship.
JSD and DJM acknowledge the support of the Royal Society through the award of a Royal Society Research Fellowship to JSD.
HMOS acknowledges support from an STFC PhD studentship and the Faculty of Science and Technology at Lancaster University.
CLH acknowledges support from the Science and Technology Facilities Council (STFC) through grant ST/Y000951/1.
CAP acknowledges the support of the UK Science
and Technology Facilities Council (STFC) via grants ST/W507441/1 and ST/Y000951/1.
This research made use of Photutils, an Astropy package for the detection and photometry of astronomical sources \citep{Bradley2024}.
This paper makes use of the following ALMA data: 
\#2011.0.00097.S,
\#2013.1.00884.S,
\#2013.1.01292.S,
\#2015.1.00055.S,
\#2015.1.00137.S,
\#2015.1.00207.S,
\#2015.1.00379.S,
\#2015.1.00664.S,
\#2015.1.00861.S,
\#2015.1.01074.S,
\#2015.1.01495.S,
\#2016.1.00279.S,
\#2016.1.00567.S,
\#2016.1.00624.S,
\#2016.1.00646.S,
\#2016.1.00726.S,
\#2016.1.01001.S,
\#2016.1.01184.S,
\#2016.1.01454.S,
\#2017.1.00046.S,
\#2017.1.00893.S,
\#2017.A.00013.S,
\#2018.1.00231.S,
\#2018.1.00478.S,
\#2018.1.01044.S,
\#2018.1.01128.S,
\#2018.1.01225.S,
\#2018.1.01871.S,
\#2019.1.00244.S,
\#2019.1.00477.S,
\#2019.1.01142.S,
\#2019.1.01722.S,
\#2021.1.00225.S,
\#2021.1.00705.S,
\#2021.1.01005.S,
\#2021.1.01500.S,
\#2022.1.00863.S.
\#2023.A.00021.S,
ALMA is a partnership of ESO (representing its member states), NSF (USA), and NINS (Japan), together with NRC (Canada), MOST and ASIAA (Taiwan), and KASI (Republic of Korea), in cooperation with the Republic of Chile. The Joint ALMA Observatory is operated by ESO, AUI/NRAO, and NAOJ.
\end{acknowledgements}
%
%
\bibliographystyle{aa_url}
\bibliography{bib}

\onecolumn
\begin{appendix}\label{sec:appflux} 
\section{Median stacked fluxes}
\vspace{30pt}
\begin{table*}[h] 
    \centering
    \tiny
    \setlength{\tabcolsep}{3pt}
    \renewcommand{\arraystretch}{0.95}

    \vspace*{2.5cm} 

    \rotatebox{90}{%
        \begin{minipage}{\textwidth} 
            \vspace*{\fill}

            \subcaptionbox{Weighted median JWST fluxes used in STARDUST SED fitting in $\mu$Jy. Upper limits are $3\sigma$. \label{tab:JWSTflux}}{%
                \begin{tabular}{lccccccccccccc}
                \toprule
                Line & z & F090W & F115W & F150W & F200W & F277W & F356W & F410M & F444W & F460N & F470N & F770W & F1800W \\ \midrule
                Pa-$\alpha$ & 1.50 & $0.138\pm0.022$ & $0.247\pm0.026$ & $0.206\pm0.030$ & $0.168\pm0.039$ & $0.168\pm0.027$ & $0.173\pm0.026$ & $0.172\pm0.030$ & $0.176\pm0.048$ & $0.276\pm0.032$ & $0.316\pm0.060$ & $<0.438$ & $0.333 \pm 0.069$ \\
                $\mathrm{N}_{\mathrm{Pa-}\alpha}$ & 1.50 & 75 & 77 & 77 & 77 & 77 & 75 & 75 & 77 & 77 & 77 & 39 & 37 \\ \midrule
                Pa-$\beta$ & 2.65 & $0.200\pm0.027$ & $0.191\pm0.036$ & $0.315\pm0.055$ & $0.403\pm0.097$ & $0.330\pm0.080$ & $0.354\pm0.099$ & $0.339\pm0.076$ & $0.358\pm0.093$ & $0.489\pm0.072$ & $<1.432$ & $<1.804$ & $<6.783$ \\
                $\mathrm{N}_{\mathrm{Pa-}\beta}$ & 2.65 & 31 & 31 & 31 & 31 & 31 & 31 & 31 & 31 & 31 & 31 & 22 & 22 \\
                \bottomrule
                \end{tabular}
            }

            \vspace{10pt}

            \subcaptionbox{Weighted median Hubble fluxes used in STARDUST SED fitting in $\mu$Jy and the respective number of sources with available fluxes. Upper limits are $3\sigma$. \label{tab:Hubbleflux}}{%
                \begin{tabular}{lcccccccc}
                \toprule
                Line & z & F275W & F435W & F606W & F814W & F125W & F140W & F160W \\ \midrule
                Pa-$\alpha$ & 1.50 & $0.038\pm0.007$ & $0.096\pm0.022$ & $0.101\pm0.018$ & $0.113\pm0.012$ & $0.242\pm0.034$ & $0.227\pm0.062$ & $0.248\pm0.030$ \\
                $\mathrm{N}_{\mathrm{Pa-}\alpha}$ & 1.50 & 66 & 66 & 77 & 77 & 77 & 73 & 77 \\ \midrule
                Pa-$\beta$ & 2.65 & $<0.042$ & $0.107\pm0.032$ & $0.187\pm0.049$ & $0.163\pm0.040$ & $0.199\pm0.041$ & $0.252\pm0.037$ & $0.322\pm0.074$ \\
                $\mathrm{N}_{\mathrm{Pa-}\beta}$ & 2.65 & 21 & 30 & 31 & 31 & 31 & 31 & 31 \\
                \bottomrule
                \end{tabular}
            }

            \vspace{10pt}

            \subcaptionbox{Weighted median Mid–FIR fluxes used in STARDUST SED fitting in mJy. Upper limits are $3\sigma$. All sources were covered in these bands. \label{tab:FIRflux}}{%
                \begin{tabular}{lccccccccccc}
                \toprule
                Line & z & MIPS-24$\mu$m & MIPS-70$\mu$m & MIPS-160$\mu$m & Pacs-100$\mu$m & Pacs-160$\mu$m & Herschel-250$\mu$m & Herschel-350$\mu$m & Herschel-500$\mu$m & SCUBA2-850$\mu$m & AzTec-1.1 mm \\ \midrule
                Pa-$\alpha$ & 1.50 & $<4.578\times10^{-4}$ & $<0.035$ & $<0.176$ & $<0.155$ & $<3.165$ & $<3.479$ & $<2.814$ & $<3.419$ & $<0.357$ & $<1.054$ \\
                Pa-$\beta$ & 2.65 & $<4.510\times10^{-4}$ & $<0.044$ & $<0.324$ & $<0.451$ & $<3.492$ & $<3.545$ & $<4.859$ & $<6.276$ & $<0.447$ & $<1.788$ \\
                \bottomrule
                \end{tabular}
            }

            \vspace{10pt}

            \subcaptionbox{Weighted median ALMA fluxes used in STARDUST SED fitting in $\mu$Jy. Upper limits are $3\sigma$. Entries with * are highly biased. \label{tab:ALMAflux}}{%
                \begin{tabular}{lccccc}
                \toprule
                Line & z & Band 3 & Band 4 & Band 6 & Band 7 \\ \midrule
                ${\mathrm{Pa-}\alpha}$ & 1.50 & $<1.744$ & $<31.043$ & $<16.587$ & $<41.217$ \\
                $\mathrm{N}_{\mathrm{Pa-}\alpha}$ & 1.50 & 38 & 77 & 24 & 24 \\ \midrule
                ${\mathrm{Pa-}\beta}$ & 2.65 & $<3.784$ & $<37.244$ & $<91.246^*$ & $<562.699^*$ \\
                $\mathrm{N}_{\mathrm{Pa-}\beta}$ & 2.65 & 17 & 31 & 3 & 5 \\
                \bottomrule
                \end{tabular}
            }

            \vspace*{\fill} 
        \end{minipage}%
    }

    \caption{All weighted median photometry tables (a)--(d).}
    \label{tab:allflux}
\end{table*}
\vspace*{-30pt}
\newpage

\section{Mean stacked fluxes}
\vspace{35pt}
\begin{table*}[h] 
    \centering
    \tiny
    \setlength{\tabcolsep}{3pt}
    \renewcommand{\arraystretch}{0.95}

    \vspace*{3cm} 

    \rotatebox{90}{%
        \begin{minipage}{\textwidth} 
            \vspace*{\fill}

            \subcaptionbox{Weighted mean JWST fluxes in $\mu$Jy. Upper limits are $3\sigma$. \label{tab:meanJWSTflux}}{%
                \begin{tabular}{lccccccccccccc}
                \toprule
                Line & z & F090W & F115W & F150W & F200W & F277W & F356W & F410M & F444W & F460N & F470N & F770W & F1800W \\ \midrule
                Pa-$\alpha$ & 1.50 & $0.271 \pm 0.042$ & $0.464 \pm 0.066$ & $0.478 \pm 0.074$ & $0.437 \pm 0.076$ & $0.452 \pm 0.071$ & $0.47 \pm 0.077$ & $0.54 \pm 0.095$ & $0.584 \pm 0.103$ & $0.732 \pm 0.135$ & $0.899 \pm 0.182$ & $<1.476$ & $<8.111$ \\
                $\mathrm{N}_{\mathrm{Pa-}\alpha}$ & 1.50 & 75 & 77 & 77 & 77 & 77 & 75 & 75 & 77 & 77 & 77 & 39 & 37 \\ \midrule 
                Pa-$\beta$ & 2.65 & $0.259 \pm 0.044$ & $0.298 \pm 0.054$ & $0.465 \pm 0.080$ & $0.707 \pm 0.145$ & $0.554 \pm 0.102$ & $0.626 \pm 0.122$ & $0.713 \pm 0.142$ & $0.766 \pm 0.160$ & $1.050 \pm 0.261$ & $0.983 \pm 0.220$ & $<4.360$ & $<13.277$ \\
                $\mathrm{N}_{\mathrm{Pa-}\beta}$ & 2.65 & 31 & 31 & 31 & 31 & 31 & 31 & 31 & 31 & 31 & 31 & 22 & 22 \\
                \bottomrule
                \end{tabular}
            }

            \vspace{10pt}

            \subcaptionbox{Weighted median Hubble fluxes used in STARDUST SED fitting in $\mu$Jy and the respective number of sources with available fluxes. Upper limits are $3\sigma$. \label{tab:meanHubbleflux}}{%
                \begin{tabular}{lcccccccc}
                \toprule
                Line & z & F275W & F435W & F606W & F814W & F125W & F140W & F160W \\ \midrule
                Pa-$\alpha$ & 1.50 & $0.061 \pm 0.011$ & $0.151 \pm 0.029$ & $0.174 \pm 0.029$ & $0.232 \pm 0.039$ & $0.518 \pm 0.081$ & $0.562 \pm 0.111$ & $0.595 \pm 0.103$ \\
                $\mathrm{N}_{\mathrm{Pa-}\alpha}$ & 1.50 & 66 & 66 & 77 & 77 & 77 & 73 & 77 \\ \midrule
                Pa-$\beta$ & 2.65 & $<0.026$ & $0.110 \pm 0.022$ & $0.241 \pm 0.043$ & $0.261 \pm 0.057$ & $0.373 \pm 0.081$ & $0.443 \pm 0.108$ & $0.502 \pm 0.127$ \\
                $\mathrm{N}_{\mathrm{Pa-}\beta}$ & 2.65 & 21 & 30 & 31 & 31 & 31 & 31 & 31 \\
                \bottomrule
                \end{tabular}
            }

            \vspace{10pt}

            \subcaptionbox{Weighted mean Mid–FIR fluxes in mJy. Upper limits are $3\sigma$. All sources were covered in these bands. \label{tab:meanFIRflux}}{%
                \begin{tabular}{lccccccccccc}
                \toprule
                Line & z & MIPS-24$\mu$m & MIPS-70$\mu$m & MIPS-160$\mu$m & Pacs-100$\mu$m & Pacs-160$\mu$m & Herschel-250$\mu$m & Herschel-350$\mu$m & Herschel-500$\mu$m & SCUBA2-850$\mu$m & AzTec-1.1 mm \\ \midrule
                Pa-$\alpha$ & 1.50 & $<6.504\times10^{-4}$ & $<0.035$ & $<0.194$ & $<0.189$ & $<4.522$ & $<3.551$ & $<3.164$ & $<4.323$ & $<0.291$ & $<0.658$ \\
                Pa-$\beta$ & 2.65 & $<7.782\times10^{-4}$ & $<0.042$ & $<0.451$ & $<1.437$ & $<7.219$ & $<2.968$ & $<4.612$ & $<5.503$ & $<0.479$ & $<1.704$ \\
                \bottomrule
                \end{tabular}
            }

            \vspace{10pt}

            \subcaptionbox{Weighted mean ALMA fluxes in $\mu$Jy. Upper limits are $3\sigma$. Entries with * are highly biased. \label{tab:meanALMAflux}}{%
                \begin{tabular}{lccccc}
                \toprule
                Line & z & Band 3 & Band 4 & Band 6 & Band 7 \\ \midrule
                ${\mathrm{Pa-}\alpha}$ & 1.50 & $<1.633$ & $<15.671$ & $<16.221$ & $<53.923$ \\
                $\mathrm{N}_{\mathrm{Pa-}\alpha}$ & 1.50 & 38 & 77 & 24 & 24 \\ \midrule
                ${\mathrm{Pa-}\beta}$ & 2.65 & $<2.910$ & $<31.823$ & $30.919 \pm 4.670^*$ & $<518.793^*$ \\
                $\mathrm{N}_{\mathrm{Pa-}\beta}$ & 2.65 & 17 & 31 & 3 & 5 \\
                \bottomrule
                \end{tabular}
            }

            \vspace*{\fill} 
        \end{minipage}%
    }

    \caption{All weighted mean photometry tables (a)--(d).}
    \label{tab:meanallflux}
\end{table*}
\vspace*{-30pt}
\newpage

\section{Plots of relations between SFR$_{\rm 100\,Myr}$, SFR$_{\rm Line}$, and $A_{\rm Line}$}
\begin{figure*}[hb]
    \sidecaption
    \includegraphics[width=6cm]{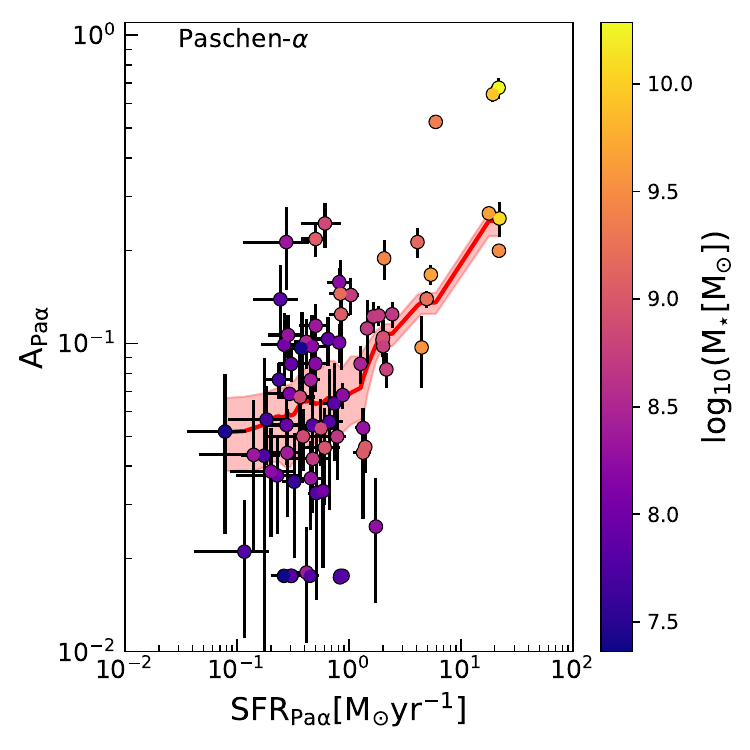}
    \includegraphics[width=6cm]{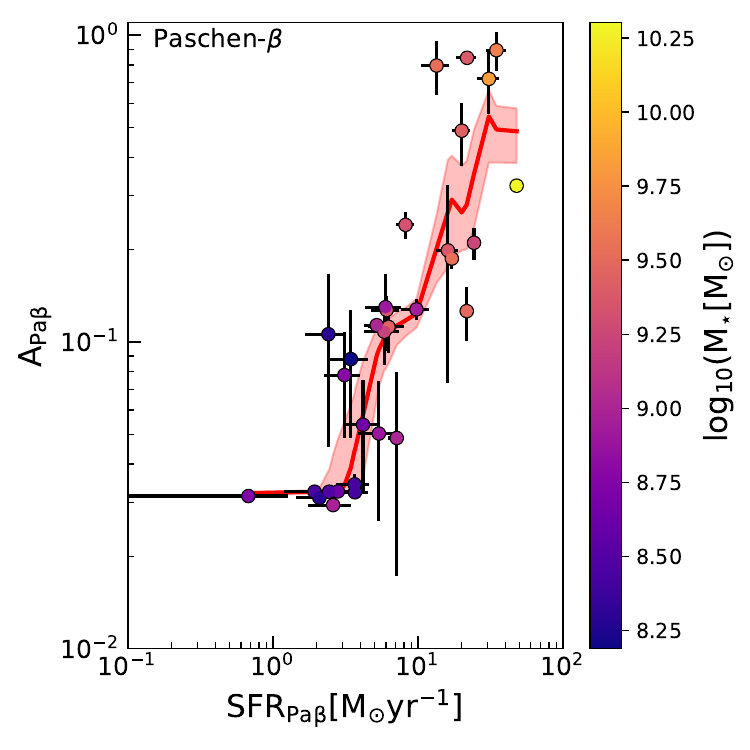}
    \caption{Relationships between $A_{\rm Line}$ and $\rm SFR_{Line}$, for the Pa-$\alpha$ (\textit{left}) and Pa-$\beta$ (\textit{right}) samples. The points are colored by their stellar mass. $A_{\rm Line}$ is estimated from converting the $A_{\rm V,\star}$ value from SED fitting using a \cite{Cardelli1989} curve.}
    \label{fig:ALine2}
\end{figure*}
\begin{figure*}[hb]
    \sidecaption
    \includegraphics[width=6cm]{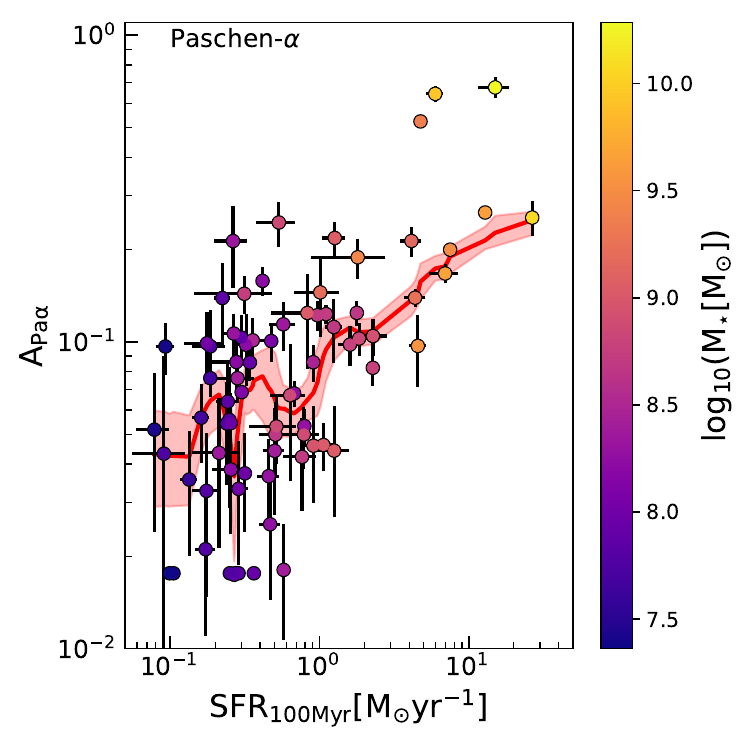}
    \includegraphics[width=6cm]{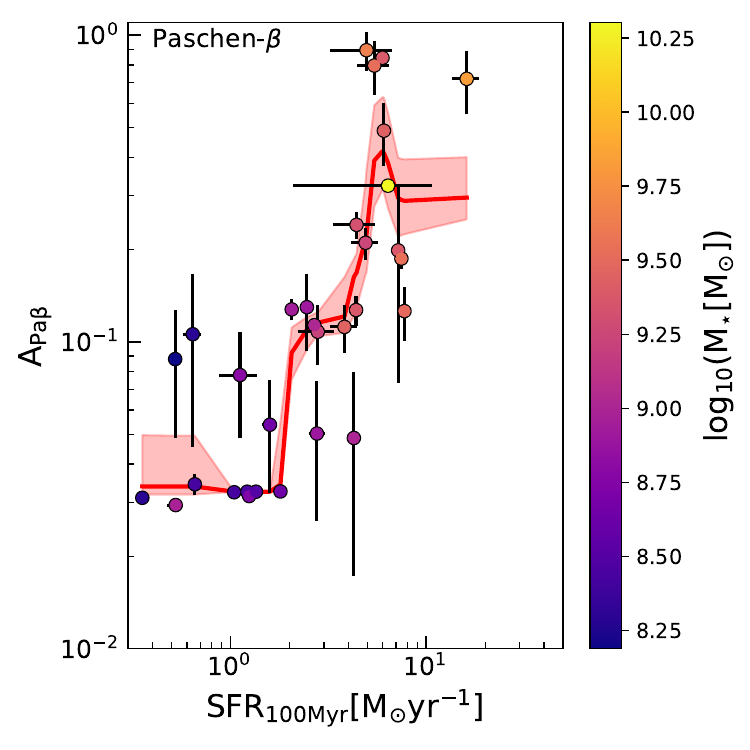}
    \caption{Relationships between $A_{\rm Line}$ and SFR$\rm_{100\, Myr}$ from SED fitting, for the Pa-$\alpha$ (\textit{left}) and Pa-$\beta$  (\textit{right}) samples. The points are colored by their stellar mass. $A_{\rm Line}$ is estimated from converting the $A_{\rm V,\star}$ value from SED fitting using a \cite{Cardelli1989} curve.}
    \label{fig:ALine3}
\end{figure*}

\section{Weighted median and mean stacks of ALMA Band 4 for Paschen-$\beta$ emitters}
\begin{figure*}[hb]
    \sidecaption
    \includegraphics[width=6cm]{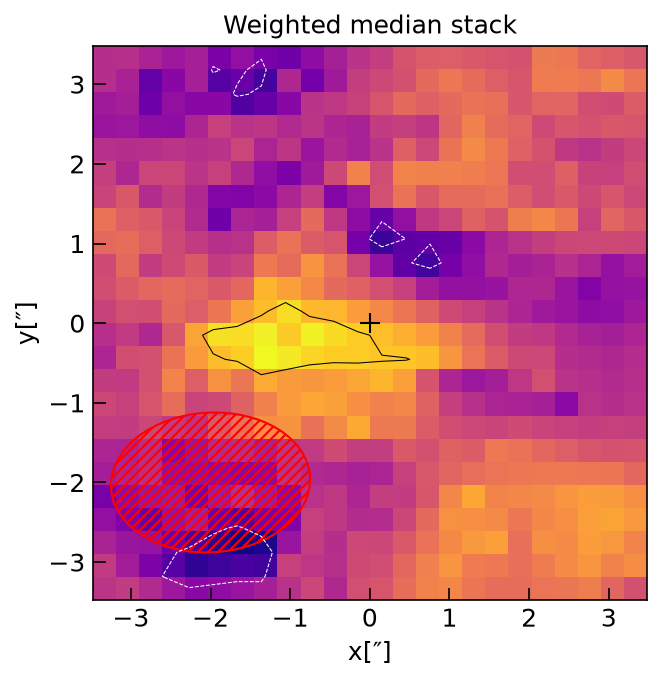}
    \includegraphics[width=6cm]{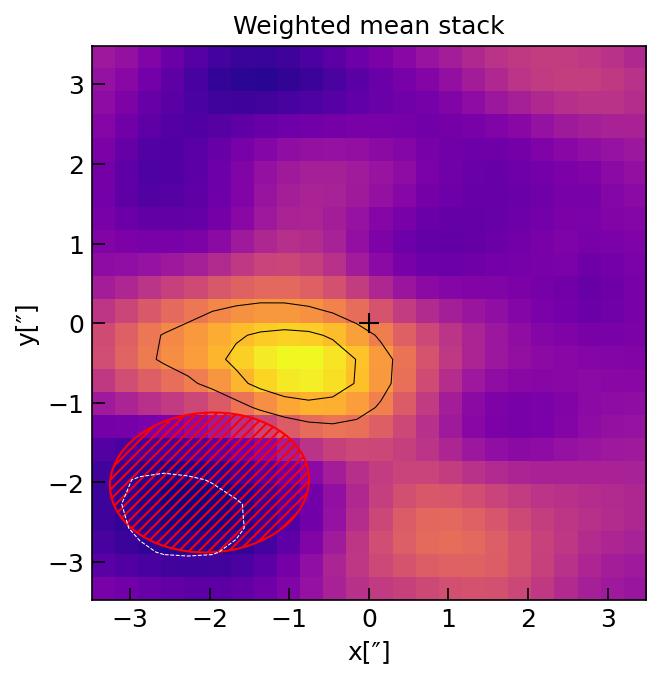}
    \caption{The $7''\times7''$ median (\textit{left}) and mean (\textit{right}) stacked images for the Paschen-$\beta$ sample using ALMA Band 4 data. The $\sigma$ contours are calculated by taking the $3\sigma$ sigma-clipped standard deviation of the image pixels. The contours start at $2\sigma$ and increase in steps of $1\sigma$. The beam is shown with a red hatched ellipse in the corner.}
    \label{fig:Band4stacks}
\end{figure*}
\section{Gas mass estimates}\label{sec:Gas mass}
To ascertain an upper limit for the gas mass, we have to convert the FIR measurement into an estimate of the gas mass, which requires a dust mass estimate and a dust-to-gas mass ratio. 
We use Band 4 to estimate an upper limit for the dust mass. We put constraints on the available dust mass by assuming a dust mass absorption coefficient of 0.077\,m$^2$\,kg$^{-1}$ \citep{Dunne2000}, the T$_{\rm dust}$ from the previous equation, and the same gray body spectrum, we derive 3$\sigma$ upper limits for Pa-$\alpha$ of $M_{\rm dust}\lesssim10^{8.04}\,\rm M_{\odot}$ and for Pa-$\beta$ of $M_{\rm dust}\lesssim10^{7.94}\,\rm M_{\odot}$.
Previous $S_{\rm 850\,\mu m}$ to $M_{\rm gas}$ 
calibrations have primarily been derived from massive ($>10^{9.5}\,\rm M_{\odot}$), near-solar-metallicity galaxies at low redshift and at $z\sim2$, effectively assuming Milky Way–like dust abundances and a gas-to-dust ratio (GDR) of $\approx100$ for these metal-rich systems
\citep[e.g.,][]{Scoville2013, Genzel2015, Scoville2016, Dunne2022}. 
In contrast, our galaxy samples generally have lower stellar masses and are expected to have higher GDRs.
Utilizing Eq. 5 from \cite{Popping2023} relating the metallicity to the GDR, which was derived from fitting galaxies in the range $\rm 8.4 < 12 + \log (O/H) < 8.8$ together with the results from the Multi-Object Spectrometer
For Infrared Exploration survey \citep[MOSDEF;][]{Sanders2021} showing that stacked galaxies at $z\sim2.3$ 
with $M_\star=10^{8.60}\,\rm M_{\odot}$ have $12 + \log (O/H)\sim7.80$ by extrapolation and for 
$M_\star=10^{9.15}\,\rm M_{\odot}$ have $12 + \log (O/H)\sim8.25$, we obtain an estimate of the ${\rm GDR}\sim1740$ for Pa-$\alpha$ and ${\rm GDR}\sim475$ for Pa-$\beta$. We note that similar results are obtained by using the relation of \cite{Ruyer2014}, which is calibrated for $z=0$ galaxies.
These estimates must be viewed in the context of the relatively large error associated with the \cite{Popping2023} equation and the fact that we extrapolate beyond the range of the fit. 

Considering the previous $M_{\rm dust}$ values and assuming these GDRs, we estimate a 3$\sigma$ upper limit for the gas mass of $M_{\rm gas}\lesssim10^{11.24}\,\rm M_{\odot}$ for the Pa-$\alpha$ and $M_{\rm gas}\lesssim10^{10.60}\,\rm M_{\odot}$ for the Pa-$\beta$ sample. Given that the stellar mass from the STARDUST fits are $10^{8.60}\, M_{\odot}$ and $10^{9.15}\,\rm M_{\odot}$ for the two samples. 
This only places very loose constraints on the median stellar mass-to-gas mass ratio, too weak to determine the typical gas reservoir of these galaxies and their gas depletion time. It is clear that deeper observations of these Paschen line emitters are required to constrain their dust continuum emission and, consequently, their gas mass. Deep ALMA continuum observations in Band 6 or 7 would have the best chance of placing strong constraints on the dust SED, as they are closer to the expected peak dust emission. Further work also has to be conducted on the dust-to-gas ratio in low-mass galaxies beyond the local universe. Deep spectra from instruments such as NIRSpec on the JWST could help in determining the metallicity and GDR for these galaxies.
\section{Percentage of narrowband excess sources in JELS}
\begin{figure*}[h]
    \sidecaption
    \includegraphics[width=6cm]{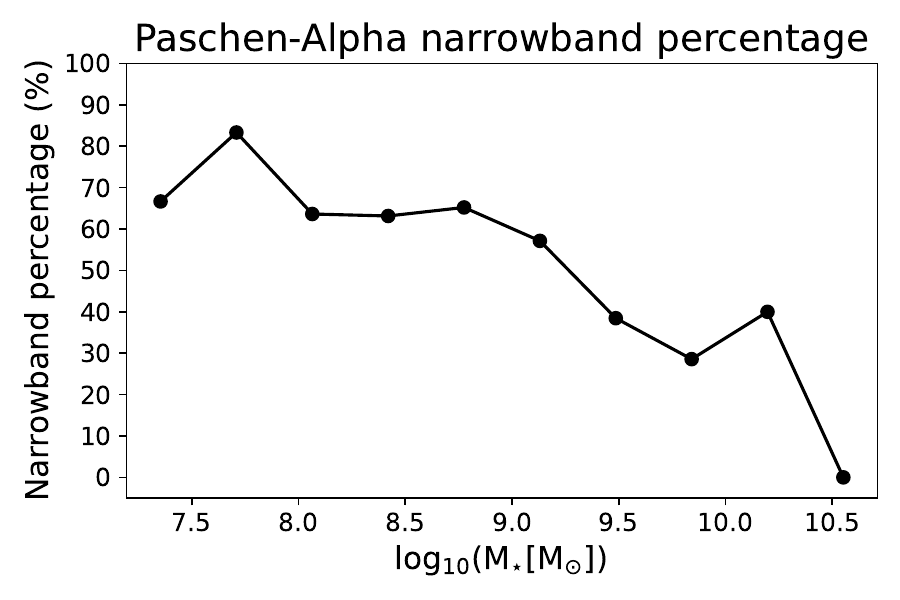}
    \includegraphics[width=6cm]{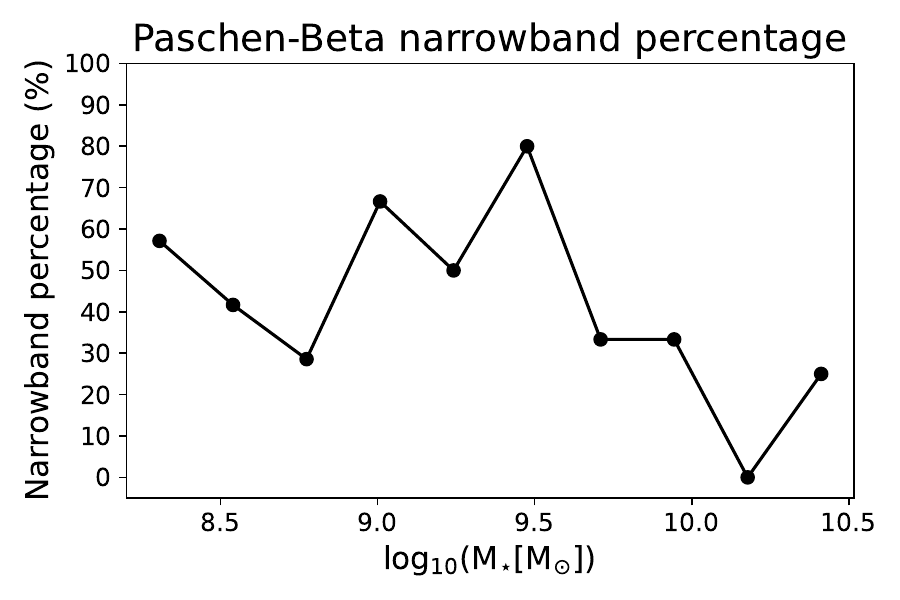}
    \caption{Percentage of narrowband excess sources compared to all sources within the narrowband redshift ranges in JELS for Paschen-$\alpha$ (\textit{left}) and Paschen-$\beta$ (\textit{right}) binned in stellar mass. Both samples are divided into ten bins.}
    \label{fig:Cratios}
\end{figure*}

\end{appendix}

\end{document}